\documentclass[fleqn,usenatbib,usedcolumn]{mnras}

\usepackage[british]{babel}             
\usepackage{newtxtext}                  
   \usepackage[slantedGreek]{newtxmath}    
   \usepackage[T1]{fontenc}                
   \usepackage{graphicx}                   
\usepackage{lscape}
\usepackage{longtable}
\usepackage{tabu}
\usepackage{url}
\usepackage{breakurl}
\usepackage{bm}
\renewcommand{\vec}[1]{\boldsymbol{#1}}

\title[Physical and observational cluster modelling]{Comparison of physical and observational galaxy cluster modelling}

\author[K. Javid et al.]
{Kamran Javid$^{1,2}$\thanks{E-mail: kj316@mrao.cam.ac.uk},
Yvette C. Perrott$^{1}$,
Michael P. Hobson$^{1}$,
Malak Olamaie$^{3,1}$,
\newauthor
Clare Rumsey$^{1}$,
and Richard D. E. Saunders$^{1,2}$
\\
$^{1}$Astrophysics Group, Cavendish Laboratory, JJ Thomson Avenue, Cambridge CB3 0HE, UK\\
$^{2}$Kavli Institute for Cosmology Cambridge, Madingley Road, Cambridge, CB3 0HA, UK\\
$^{3}$Imperial Centre for Inference and Cosmology (ICIC), Imperial College, Prince Concort Road, London, SW7 2AZ, UK \\
}

\date{Accepted XXX. Received YYY; in original form ZZZ}

\pubyear{2018}

\begin{document}
\label{firstpage}
\pagerange{\pageref{firstpage}--\pageref{lastpage}}
\maketitle

\begin{abstract}
We present a comparison between three cluster models applied to data obtained by the Arcminute Microkelvin Imager radio interferometer system. The physical model (PM) parameterises a cluster in terms of its physical quantities to model the dark matter and baryonic components of the cluster using Navarro-Frenk-White (NFW) and generalised-NFW profiles respectively. The observational models (OM I and OM II) model only the gas content of the cluster. The two OMs vary only in the priors they use in Bayesian inference: OM I has a joint prior on angular radius $\theta$ and integrated Comptonisation $Y$, derived from simulations, while OM II uses separable priors on $\theta$ and $Y$ which are based on calculations of the physical model.   
For the comparison we consider a sample of $54$ clusters which are a subsample of the second \textit{Planck} catalogue of Sunyaev--Zel'dovich sources. \\
We first compare the $Y$ estimates of the three models, and find that the PM generally yields lower estimates relative to the OMs. 
We then compute the Earth Mover's Distance between the $\theta$ -- $Y$ posterior distributions obtained from each model for each cluster, and find that the two models which are most discrepant are PM and OM I. \\
Finally, we compare the Bayesian evidence values obtained from each model for each cluster. OM I generally provides the best fit to the data but not at a statistically significant level, according to the Jeffreys scale. The highest evidence ratio obtained is actually in favour of the PM over OM I.
\end{abstract}



\section{Introduction}
This paper provides a follow-up to the work presented in \citet{2019MNRAS.483.3529J} (from here on referred to as KJ19), in which we performed Bayesian inference on data obtained with the Arcminute Microkelvin Imager (AMI) radio interferometer system, to derive estimates of physical properties of clusters that have been detected by \textit{Planck}. 
In this paper we focus on the observational properties of clusters obtained from telescopes such as AMI and \textit{Planck} which measure the Sunyaev--Zel'dovich (SZ, \citealt{1970CoASP...2...66S}) effect: the angular radius, $\theta$, and the integrated Comptonisation parameter, $Y$. 
For the sample considered in KJ19, we compare observational parameters derived from the physical model based on that derived in \citet{2012MNRAS.423.1534O} (from here on MO12) with those obtained from two observational models similar to the one described in \citet{2015A&A...580A..95P} (from here on YP15) and \citet{2012MNRAS.421.1136A}, using data from AMI. 
Furthermore, we compare the different models using Bayesian analysis.

The paper is organised as follows: In Section~\ref{sec:telescopes} we give an overview of the \textit{Planck} mission and the AMI array, and how the cluster sample for the analysis was selected. In Section~\ref{sec:models} we review how the physical modelling process for data obtained from AMI. We also summarise the observational model presented in YP15, and introduce a similar model which implicitly encodes redshift information into the model through the priors. 
Section~\ref{sec:results_ii} presents the results of the Bayesian model selection analysis performed on the physical and observational models using AMI data, as well as a comparison of their posterior distributions (using a metric defined for distributions). Finally, we provide a summary and discuss future work in Section~\ref{sec:summary}.

In this work a `concordance' flat $\Lambda$CDM cosmology is assumed.


\section{\textit{Planck} and AMI telescopes, and the cluster sample}
\label{sec:telescopes}


\subsection{\textit{Planck} mission}
The combination of the \textit{Planck} satellite's low frequency and high frequency instruments provide nine frequency channels in the range 37~GHz -- 857~GHz.
Of particular importance to the cluster sample considered here are the \textit{Planck} catalogues of SZ clusters (see \citealt{2014A&A...571A..29P}, \citealt{2015A&A...581A..14P} and \citealt{2016A&A...594A..27P} for papers relating to catalogues PSZ1, PSZ1.2 and PSZ2 respectively, where `PSZX' refers to the X\textsuperscript{th} \textit{Planck} SZ catalogue). Here we use the data provided in PSZ2, as it is the most recent \textit{Planck} SZ catalogue. PSZ2 gives the sky coordinates at which AMI made observations, and the redshift ($z$) information required in the modelling (for more on the sources of the $z$ values, see Section~2.2 of KJ19).


\subsection{AMI}
AMI is a dual-array interferometer designed for SZ studies, which is situated near Cambridge, UK.  AMI consists of two arrays: the Small Array (SA), optimised for viewing arcminute-scale features, having an angular resolution of $\approx$\,3\,arcmin and sensitivity to structures up to $\approx$\,10\,arcmin in scale; and the Large Array (LA), with angular resolution of $\approx$\,30\,arcsec, which is insensitive to the arcminute-scale emission due to clusters and is used to characterise and subtract confusing radio-sources.  Both arrays operate at a central frequency of $\approx$\,15\,GHz with a bandwidth of $\approx$\,4.5\,GHz, divided into six channels.  For further details of the instrument see \citet{2008MNRAS.391.1545Z}.
Note that the AMI array has recently upgraded from an analogue correlator to a digital correlator \citep{2018MNRAS.475.5677H}, but the data used in this analysis were obtained using its analogue correlator.


\subsection{Selection of the cluster sample}
Based on AMI's observational capability, and values taken from the PSZ2 catalogue, the initial cluster selection in KJ19 was based on the following: 
\begin{itemize}
\item The observation declination limits for AMI were set to $20^{\circ} < \delta < 87^{\circ}$ to adhere to AMI's `easy' observing limits.
\item There were no restrictions on the values of redshift taken from the \textit{Planck} catalogue.
\item The minimum \textit{Planck} signal-to-noise ratio (S/N) value for which an observation with AMI would be made was $4.5$.
\item The automatic radio-source environment rejection used in YP15 was also used in KJ19.
\end{itemize}
This led to an initial cluster sample size of 199, which had been detected by \textit{Planck} and re-observed with AMI to produce data which could be run through the data analysis pipeline. 
After posterior distributions for 197 of these clusters were produced, the sample size was reduced further as follows:
\begin{itemize}
\item Of the 197 clusters for which posterior distributions could be inferred, 73 showed good constraints on the cluster mass.
\item Seven of the 73 well constrained datasets were rejected after manual radio-source environment inspection, leaving a sample size of 66.
\item A further seven clusters were discarded due to ambiguity in their cluster centre, which rendered their parameter estimates unreliable. This left a penultimate sample size of 59.
\item Finally, five clusters were discarded due to the fact that they were not detected by the \textit{Planck} detection algorithm, PowellSnakes (PwS, \citealt{2012MNRAS.427.1384C}). These were discarded from the sample of 59, as in KJ19 only clusters with data from PwS were analysed. 
\end{itemize}
We choose to focus on the 54 cluster sample in this work, as the methodology and justification for discarding clusters mentioned above are equally applicable when considering the parameter estimation and model comparison presented here.
The maximum and minimum values of some key parameters for this sample from PSZ2 are given in Table~\ref{tab:initial_sample}. 

\begin{table}
\centering
\begin{tabular}{{l}{c}{c}}
\hline
Parameter & Min. value & Max. value \\
\hline 
Declination & $20.31^{\circ}$ & $78.39^{\circ}$ \\
$z$ & $0.0894$ & $0.83$ \\
S/N & 4.97 & 28.40 \\
$Y_{\rm marg}(5r_{500})$~($\times 10^{-3}~\rm{arcmin}^{2}$) & $0.85$ & $33.6$ \\
\hline
\end{tabular}
\caption{Minimum and maximum values for a selection of parameters taken from \textit{Planck} catalogue for the AMI sample of 54 clusters. $Y_{\rm marg}(5r_{500})$ refers to the integrated Comptonisation parameter up to a radius $5 \times r_{500}$ as discussed in Section~3.3 of KJ19.}\label{tab:initial_sample}
\end{table}



\section{Modelling AMI data}
\label{sec:models}

Our AMI Bayesian data analysis pipeline, \textsc{McAdam} closely resembles the one described in \citet{2009MNRAS.398.2049F} (FF09 from here on), but with different cluster models.
Here three different models are applied to AMI data to obtain estimates for observational parameters. 


\subsection{Bayesian inference}

Our analysis of AMI data is built upon the principles of Bayesian inference. We now give a summary of this framework in the context of both parameter estimation and model comparison.

\subsubsection{Parameter estimation}

Given a model $\mathcal{M}$ and data $\vec{\mathcal{D}}$ we can obtain the model parameter probability distributions (also known as input parameters or sampling parameters) $\vec{\Theta}$ conditioned on $\mathcal{M}$ and $\vec{\mathcal{D}}$ using Bayes' theorem:
\begin{equation}\label{eqn:bayes}
Pr\left(\vec{\Theta}|\vec{\mathcal{D}},\mathcal{M}\right) = \frac{Pr\left(\vec{\mathcal{D}}|\vec{\Theta},\mathcal{M}\right)Pr\left(\vec{\Theta}|\mathcal{M}\right)}{Pr\left(\vec{\mathcal{D}}|\mathcal{M}\right)},
\end{equation}
where $Pr\left(\vec{\Theta}|\vec{\mathcal{D}},\mathcal{M}\right) \equiv \mathcal{P}\left(\vec{\Theta}\right)$ is the posterior distribution of the model parameter set, $Pr\left(\vec{\mathcal{D}}|\vec{\Theta},\mathcal{M}\right) \equiv \mathcal{L}\left(\vec{\Theta}\right)$ is the likelihood function for the data, $Pr\left(\vec{\Theta}|\mathcal{M}\right) \equiv \pi\left(\vec{\Theta}\right)$ is the prior probability distribution for the model parameter set, and $Pr\left(\vec{\mathcal{D}}|\mathcal{M}\right) \equiv \mathcal{Z}$ is the Bayesian evidence of the data given a model $\mathcal{M}$. The evidence can be interpreted as the factor required to normalise the posterior over the model parameter space:
\begin{equation}\label{eqn:evidence}
\mathcal{Z}\left(\vec{\mathcal{D}}\right) = \int \mathcal{L}\left(\vec{\Theta}\right) \pi\left(\vec{\Theta}\right)\, \mathrm{d}\vec{\Theta},
\end{equation} 
where the integral is carried out over the $N$-dimensional parameter space. For the models using AMI data considered here, the input parameters can be split into two subsets, (which are assumed to be independent of one another): cluster parameters, $\vec{\Theta}_{\rm cl}$ and radio-source or `nuisance' parameters, $\vec{\Theta}_{\rm rs}$. The sets of cluster parameters (and their respective prior distributions) required for the three models will be given in the following Sections relevant to that model. For more details on the radio-source modelling, please refer to Section~5.2 of FF09 and Section~3.2.2 of KJ19. For more information on the likelihood functions and covariance matrices used in the AMI analysis, we refer the reader to \citet{2002MNRAS.334..569H} and Sections~5.3 of FF09 and~3.2.3 of KJ19.

\subsubsection{Model comparison}
\label{subsubsec:bayesmodel}
While it is the posterior distribution which gives the model parameter estimates from the prior information and data, it is $\mathcal{Z}\left(\vec{\mathcal{D}}\right)$ which is crucial to performing model selection. The nested sampling algorithm, \textsc{MultiNest} \citep{2009MNRAS.398.1601F} is a Monte Carlo algorithm which calculates $\mathcal{Z}\left(\vec{\mathcal{D}}\right)$ by making use of a transformation of the $N$-dimensional evidence integral into a one-dimensional integral that is much easier to evaluate. The algorithm also produces samples from $\mathcal{P}\left(\vec{\Theta}\right)$ as a by-product, meaning that it is suitable for both the parameter estimation and model comparison aspects of this work. Comparing models in a Bayesian way can be done by considering the probability of a model conditioned on $\vec{\mathcal{D}}$, which can be calculated using Bayes' theorem
\begin{equation}\label{eqn:bayesmodel}
Pr\left(\mathcal{M}|\vec{\mathcal{D}}\right) = \frac{Pr\left(\vec{\mathcal{D}}|\mathcal{M}\right)Pr\left(\mathcal{M}\right)}{Pr\left(\vec{\mathcal{D}}\right)}.
\end{equation}
Hence for two models, $\mathcal{M}_{1}$ and $\mathcal{M}_{2}$, the ratio of the models conditioned on the same dataset is given by 
\begin{equation}\label{eqn:bayesmodelcomparison}
\frac{Pr\left(\mathcal{M}_{1}|\vec{\mathcal{D}}\right)}{Pr\left(\mathcal{M}_{2}|\vec{\mathcal{D}}\right)} = \frac{Pr\left(\vec{\mathcal{D}}|\mathcal{M}_{1}\right)Pr\left(\mathcal{M}_{1}\right)}{Pr\left(\vec{\mathcal{D}}|\mathcal{M}_{2}\right)Pr\left(\mathcal{M}_{2}\right)},
\end{equation}
where $Pr(\mathcal{M}_{2}) / Pr(\mathcal{M}_{1})$ is the a-priori probability ratio of the models. We set this to one, i.e. we place no bias towards a particular model before performing the analysis. Hence the ratio of the probabilities of the models given the data is equal to the ratio of the evidence values obtained from the respective models (we define $\mathcal{Z}_{i} \equiv Pr\left(\vec{\mathcal{D}}|\mathcal{M}_{i}\right)$). 
The evidence is simply the average of the likelihood function over the sampling parameter space, weighted by the prior distribution. This means that the evidence is larger for a model if more of its parameter space is likely and smaller for a model with large areas in its parameter space having low likelihood values. Moreover, a larger parameter space, either in the form of higher dimensionality or a larger domain results in a lower evidence value all other things being equal. 
Thus the evidence automatically implements Occam's razor: when you have two competing theories that make exactly the same predictions, the simpler one is the better. \citet{jeffreys} provides a scale for interpreting the ratio of evidences as a means of performing model comparison (Table~\ref{tab:jeffreys}). A value of $\ln (\mathcal{Z}_{1} / \mathcal{Z}_{2})$ above $5.0$ (less than $-5.0$) presents "strong evidence" in favour of model 1 (model 2). Values $ 2.5 \leq \ln (\mathcal{Z}_{1} / \mathcal{Z}_{2}) < 5.0$ ($ -5.0 < \ln (\mathcal{Z}_{1} / \mathcal{Z}_{2}) \leq -2.5 $) present "moderate evidence" in favour of model 1 (model 2). Values $ 1 \leq \ln (\mathcal{Z}_{1} / \mathcal{Z}_{2}) < 2.5$ ($ -2.5 < \ln (\mathcal{Z}_{1} / \mathcal{Z}_{2}) \leq -1 $) present "weak evidence" in favour of model 1 (model 2). Finally, values $ -1.0 < \ln (\mathcal{Z}_{1} / \mathcal{Z}_{2}) < 1.0 $ require "more information to come to a conclusion" over model preference. \\


\begin{table*}
\centering
\begin{tabular}{{l}{c}{c}}
\hline
$\ln (\mathcal{Z}_{1} / \mathcal{Z}_{2})$  & Interpretation & Probability of favoured model \\ 
\hline
$\leq 1.0$ & better data are needed & $\leq 0.75$ \\
$\leq 2.5 $ & weak evidence in favour of $\mathcal{M}_{1}$ & $0.923$ \\
$\leq 5.0$ & moderate evidence in favour of $\mathcal{M}_{1}$ & $0.993$ \\
$ > 5.0$ & strong evidence in favour of $\mathcal{M}_{1}$ & $ > 0.993 $ \\
\hline
\end{tabular}
\caption{Jeffreys scale for assessing model preferability based on the log of the evidence ratio of two models.}\label{tab:jeffreys}

\end{table*}


\subsection{A physical model for AMI cluster data}
\label{subsec:phys}
The physical model (from here on PM) introduced in MO12 uses $z$ information as well as other physical sampling parameters to derive physical properties of a galaxy cluster (i.e. mass, density, radius and temperature values). The model also calculates $Y(r_{500})$, which is the integrated Comptonisation parameter out to a radius $r_{500}$ from the cluster centre.
Note that in general the radius $r_{\Delta}$ is the radius from the centre at which the enclosed average total mass density is $\Delta$ times $\rho_{\rm crit}(z)$. The critical density is given by $\rho_{\rm crit}(z) = 3H(z)^{2}/8\pi G$ where $H(z)$ is the Hubble parameter (at the cluster redshift) and $G$ is Newton's constant. 

The model assumes an Navarro-Frenk-White (NFW) profile \citep{1995MNRAS.275..720N} for the dark matter component of a galaxy cluster
\begin{equation}\label{eqn:nfw}
\rho_{\rm dm}(r) = \frac{\rho_{\rm s}}{\left(\frac{r}{r_{\rm s}}\right)\left(1+\frac{r}{r_{\rm s}}\right)^{2}}.
\end{equation}
$\rho_{\rm dm}(r)$ is the dark matter density as a function of cluster radius $r$, $\rho_{\rm s}$ is an overall density normalisation coefficient and $r_{\rm s}$ is a characteristic radius defined by $r_{\rm s} = r_{200}/c_{200}$ where $c_{200}$ is the concentration parameter at $r_{200}$. 
Following the work of \citet{2007ApJ...668....1N}, the generalised-NFW model (GNFW) is used to parameterise the electron pressure as a function of radius $P_{\rm e}(r)$, from the cluster centre
\begin{equation}\label{eqn:gnfw}
P_{\rm e}(r) = \frac{P_{\rm ei}}{\left(\frac{r}{r_{\rm p}}\right)^{c}\left(1+\left(\frac{r}{r_{\rm p}}\right)^{a}\right)^{(b-c)/a}}.
\end{equation}
$P_{\rm ei}$ is an overall pressure normalisation factor and $r_{\rm p}$ is another characteristic radius, defined by $r_{\rm p} = r_{500}/c_{500}$. The parameters $a,\,b$ and $c$ describe the slope of the pressure profile at $ r \approx r_{\rm p}$, $r \gg r_{\rm p}$ and $r \ll r_{\rm p}$ respectively. 
The input parameters of the prior distributions are the same as in KJ19 (and are given in Table~\ref{tab:phys_priors}), as are the calculational steps including the modifications to MO12. Values for $z_{\rm Planck}$ were taken from the PSZ2 catalogue.

\begin{table}
\centering
\begin{tabular}{{l}{c}}
\hline
Parameter & Prior distribution \\ 
\hline
$x_{\rm c}$ & $\mathcal{N}(0'', 60'')$ \\
$y_{\rm c}$ & $\mathcal{N}(0'', 60'')$ \\
$z$ & $\delta(z_{\rm Planck})$ \\
$M(r_{200})$ & $\mathcal{U} [ \log (0.5\times 10^{14} M_{\mathrm{Sun}}), \log (50\times 10^{14} M_{\mathrm{Sun}})]$ \\
$f_{\rm gas}(r_{200})$ & $\mathcal{N}(0.13, 0.02)$ \\
\hline
\end{tabular}
\caption{Physical model input parameter prior distributions, where the normal distributions are parameterised by their mean and standard deviations.}\label{tab:phys_priors}
\end{table}


\subsection{Observational model I}
\label{subsec:obs_i}
Observational model I (OM I) is based on the one used in YP15. It uses the same GNFW profile (given by equation~\ref{eqn:gnfw} in the current paper) to model the gas content, but with the slope parameters used in KJ19; it takes into account only the cluster gas -- it does not explicitly model the dark matter component. It deals in \textit{angular} rather than physical sizes. Like the PM, OM I assumes spherical symmetry and the equation of state of an ideal gas. \\
The model has four cluster input parameters: the total integrated Comptonisation parameter, $Y_{\rm tot}$, $\theta_{\rm p}$ ($ = r_{\rm p} / D_{\rm A}$), $x_{\rm c}$ and $y_{\rm c}$. The priors used on $Y_{\rm tot}$ and $\theta_{\rm p}$ are the same as the `new' priors used in YP15. These were derived from the \textit{Planck} completeness simulations \citep{2014A&A...571A..29P} as follows. The simulations were produced by drawing a cluster population from the Tinker mass function \citep{2008ApJ...688..709T} and using the scaling relations in \citet{2011A&A...536A..11P} to obtain observable quantities. This cluster population was injected into the real \textit{Planck} data and a simulated union catalogue was created by running the \textit{Planck} detection pipelines on this simulated dataset. An elliptical Gaussian function was then fitted to the posterior of $Y_{\rm tot}$ and $\theta_{\rm p}$ in log space. Hence the prior has the \textit{Planck} selection function implicitly included in it. \\
For consistency, the same cluster centre priors were used in both observational models as in the PM. The priors for OM I are summarised in Table~\ref{tab:obs_i_priors}.
\begin{table}
\centering
\begin{tabular}{{l}{c}}
\hline
Parameter & Prior distribution \\ 
\hline
$x_{\rm c}$ & $\mathcal{N}(0'', 60'')$ \\
$y_{\rm c}$ & $\mathcal{N}(0'', 60'')$ \\
$\log (Y_{\rm tot}), \, \log(\theta_{\rm p})$ & $\mathcal{N}((-2.7, 0.62), (0.29, 0.12), 40.2^{\circ})$ \\
\hline
\end{tabular}
\caption{Observational model I input parameter prior distributions. Note that the Gaussian elliptical function on $\log  (Y_{\rm tot}) - \log (\theta_{\rm p})$ is parameterised in terms of the mean in both dimensions, the respective standard deviations and the offset of the principle axes from the vertical and horizontal axes measured clockwise.}\label{tab:obs_i_priors}
\end{table}
From $Y_{\rm tot}$ and $\theta_{\rm p}$, the observational model calculates the modelled data required for use in inference with interferometer SZ data (see FF09 Sections~4 and~5) i.e. of the same form as the physical model.


\subsection{Observational model II}
\label{subsec:obs_ii}

OM II takes the same form as OM I but the priors assigned to $Y_{\rm tot}$ and $\theta_{\rm p}$ are different: they incorporate the spectroscopic or photometric redshift of each cluster. 

From the $z$ and $M(r_{200})$ priors of the PM and for $f_{\rm{gas}}(r_{200}) = 0.13$, upper and lower bounds on $Y_{\rm tot}$ and $\theta_{\rm p}$ are calculated using the PM.
Note that $Y_{\rm tot}$ and $\theta_{\rm p}$ are assumed to be a-priori \textit{uncorrelated}, unlike in OM I. For the lowest redshift cluster ($z = 0.0894$), these limits are $ \theta_{\rm p,\, min} = 4.24~\rm{arcmin}$, $ \theta_{\rm p,\, max} = 19.04~\rm{arcmin}$, $ Y_{\rm tot,\, min} =  1.06 \times 10^{-4}~\rm{arcmin}^{2}$ and $ Y_{\rm tot,\, max} = 0.19~\rm{arcmin}^{2}$; for the highest redshift ($z = 0.83$) cluster these limits are $ \theta_{\rm p,\, min} = 0.67~\rm{arcmin}$, $ \theta_{\rm p,\, max} = 3.01~\rm{arcmin}$, $ Y_{\rm tot,\, min} =  5.7 \times 10^{-6}~\rm{arcmin}^{2}$ and $ Y_{\rm tot,\, max} = 0.01~\rm{arcmin}^{2}$. It clear that $z$ has a large effect on the PM calculations, as it is used to calculate the angular scale from $r$ through $\theta = r / D_{\rm A}(z)$ where $D_{\rm A}(z)$ is the angular diameter distance of the cluster at redshift $z$, and to convert the units of $Y$ (see Section~\ref{sec:results_ii}). It is also used to calculate $c_{200}$ which affects the scale of the self-similar dark matter density profile, and the normalisation constant $\rho_{\rm{s}}$ in equation~\ref{eqn:nfw} is proportional to $\rho_{\rm{crit}}(z)$.
The priors for OM II are summarised in Table~\ref{tab:obs_ii_priors}.
\begin{table}
\centering
\begin{tabular}{{l}{c}}
\hline
Parameter & Prior distribution \\ 
\hline
$x_{\rm c}$ & $\mathcal{N}(0'', 60'')$ \\
$y_{\rm c}$ & $\mathcal{N}(0'', 60'')$ \\
$\theta_{\rm p}$ & $\mathcal{U} [ \log ( \theta _{\rm p,\, min}(z) ),\log ( \theta _{\rm p,\, max}(z) )]$ \\
$Y_{\rm tot}$ & $\mathcal{U} [ \log ( Y_{\rm tot, \, min}(z) ), \log ( Y_{\rm tot,\, max}(z) )]$ \\
\hline
\end{tabular}
\caption{Observational model II input parameter prior distributions.}\label{tab:obs_ii_priors}
\end{table}
Note that in using these PM calculations to calculate the prior limits, we have made the assumptions underlying the PM but to which the observational model is not subject to (i.e. hydrostatic equilibrium up to radius $r_{200}$ and $f_{\rm{gas}}$ is much less than unity up to the same radius).

\section{AMI model comparisons}
\label{sec:results_ii}

We now use AMI data to compare the PM, OM I and OM II. We begin by comparing their observational parameter estimates.
Secondly we introduce a metric which measures the `distance' between probability distributions. In this context the distance is measured between the $\left(Y(r_{500}), \,\theta_{500}\right)$ posterior distributions of the three models. Finally the models are compared using the evidence ratios introduced in Section~\ref{subsubsec:bayesmodel}.
The results obtained from these analyses are given in Appendix~\ref{sec:results_tables}, which lists the values obtained for the 54 cluster sample in ascending order of $z$. \\
We emphasise the notation used for $Y$. For consistency we parameterise $Y$ by $r$ for all three models ($Y \equiv Y(r)$). For the PM, $Y(r)$ has units [length$^{2}$]; to convert this to the more conventional [angle$^{2}$] we divide by $D_{A}^{2}$: $ Y(r) \rightarrow Y(r)/D_{A}^{2} $. The $Y$ value given by an OM is naturally in units of [angle$^{2}$]; when we refer to $Y(r)$ in the context of the OM we equivalently mean $Y(\theta)$.


\subsection{Physical and observational models Y values comparison}
\label{subsec:amiycomparison}


\begin{figure*} 
  \begin{center}
  \includegraphics[ width=0.90\linewidth]{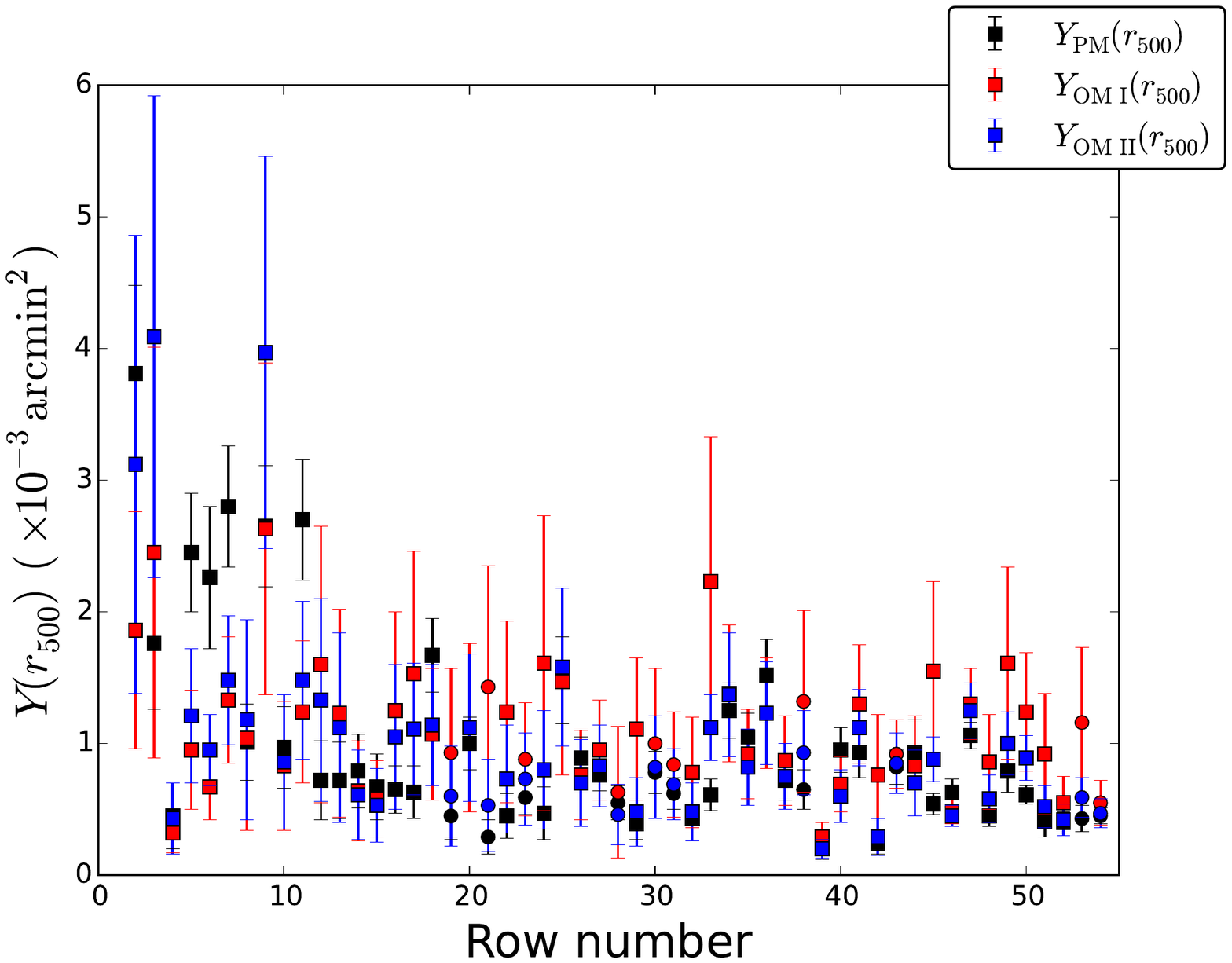}
  \caption{Plot of $Y(r_{500})$ obtained from AMI data using the physical and observational models vs row number of Table~\ref{tab:results1}. The points with circular markers correspond to clusters whose redshifts were measured photometrically as opposed to spectroscopically. For clarity purposes the first row is not plotted due to its relatively large value ($Y(r_{500}) \approx 10~\rm{arcmin}^{2}$).}
\label{graph:y500ami}
  \end{center}
\end{figure*}


\begin{figure*} 
  \begin{center}
  \includegraphics[ width=0.90\linewidth]{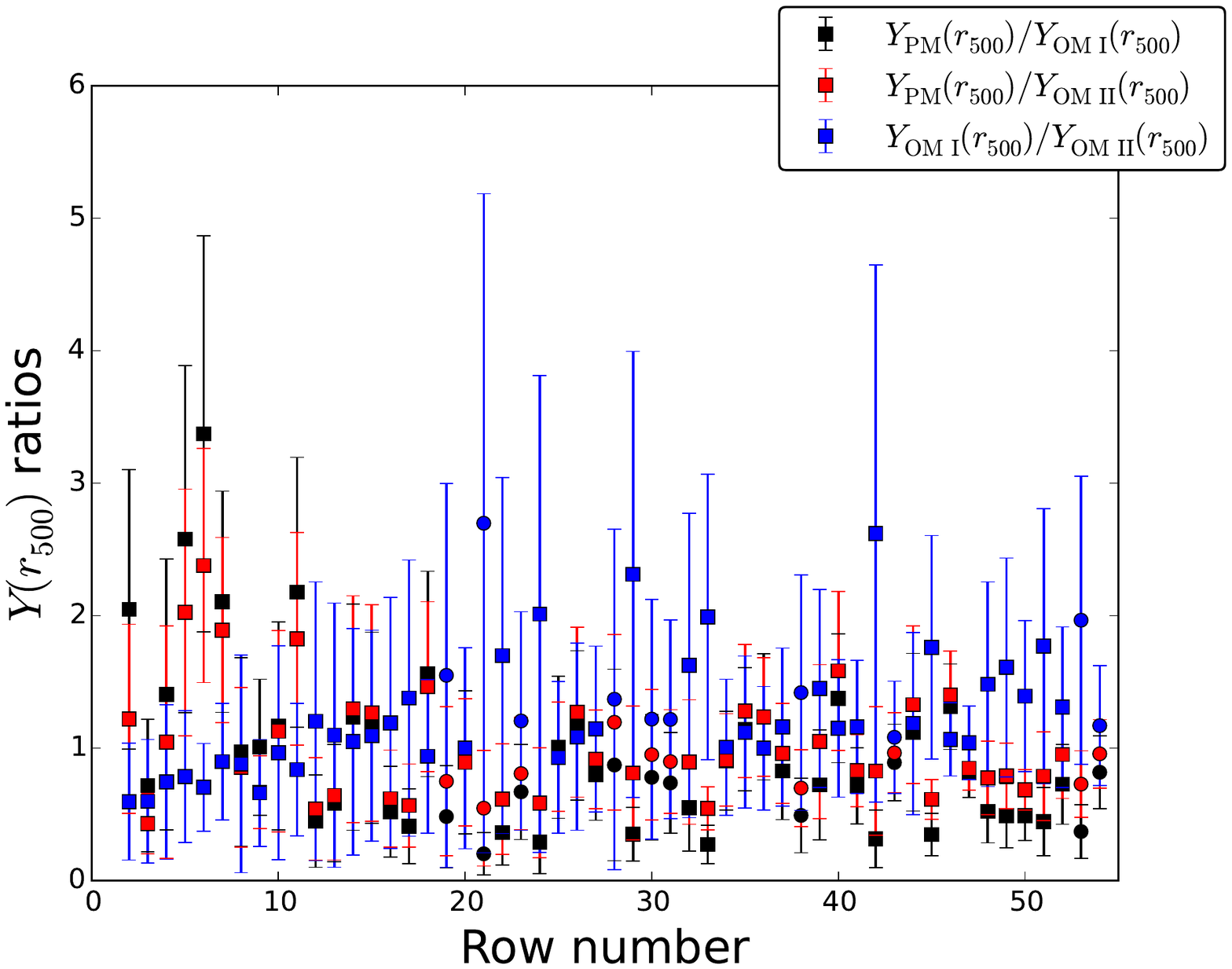}
  \caption{Plot of $Y(r_{500})$ ratio vs row number of Table~\ref{tab:results1} for three different cases: $Y_{\rm PM}(r_{500}) / Y_{\rm OM~I}(r_{500})$; $Y_{\rm PM}(r_{500}) / Y_{\rm OM~II}(r_{500})$ and  $Y_{\rm OM~I}(r_{500}) / Y_{\rm OM~II}(r_{500})$. The points with square markers correspond to clusters whose redshifts were measured spectroscopically, and the circular markers photometrically (as listed in Table~\ref{tab:results1}).}
\label{graph:y500amifrac}
  \end{center}
\end{figure*}

Figure~\ref{graph:y500ami} shows the posterior mean values for $Y(r_{500})$ for the three models used on the same AMI datasets. We first note that the errors associated with the OM estimates are generally larger than those with the PM. Secondly it appears that the OM I $Y$ values are less strongly correlated with $z$ than those from the PM and OM II. This may be because OM I contains no explicit $z$-information, and in fact its only reliance on $z$ is from the simulated and empirical datasets used to fit its prior distribution, but the same prior is used for all clusters, and so the dependence on redshift is very weak. \\ 
We now compare the results from the three models pairwise. Note that when we refer to the dispersion between values in units of standard deviations, 
 we are referring to the combined standard deviation of the two $Y$ values.
When comparing PM and OM I values of $Y$, just 15 clusters are within one standard deviation, 27 within two and 18 are more than three standard deviations away from each other. The same comparison between PM and OM II gives corresponding values of 23, 40 and 5. This implies that the dispersion between OM II and PM is much smaller (especially in the extreme cases), and shows the importance in the choice of priors. 
Table~\ref{tab:physdispersion} gives a summary of the dispersion of the PM with respect to the OMs.
Figure~\ref{graph:y500amifrac} shows the fractional difference between the $Y$ values for the three models, and shows that the PM estimates are generally much higher than both OM values at low $z$. However, in general the PM yields lower estimates $Y$ compared to the OMs (PM underestimates $Y$ relative to OM I and OM II 35 and 36 times respectively).

\begin{table*}
\centering
\begin{tabular}{{l}{c}{c}{c}}
\hline
Model comparison ($Y_{\mathcal{M}_{i}} \equiv$) & $|Y_{\rm{PM}} - Y_{\mathcal{M}_{i}}| / \sigma_{\rm{PMand} \mathcal{M}_{i}} < 1$ & $|Y_{\rm{PM}} - Y_{\mathcal{M}_{i}}| / \sigma_{\rm{PMand} \mathcal{M}_{i}} < 2$ & $|Y_{\rm{PM}} - Y_{\mathcal{M}_{i}}| / \sigma_{\rm{PMand} \mathcal{M}_{i}} > 3$ \\ 
\hline
$Y_{\rm OM \, I}$ & $15$ & $27$ & $18$ \\
$Y_{\rm OM \, II}$ & $23$ & $40$ & $5$ \\
\hline
\end{tabular}
\caption{Difference between physical model mean values for $Y(r_{500})$ and observational model mean values, measured in units of the physical model $Y(r_{500})$ standard deviation. The numbers in the columns correspond to the number of clusters out of the sample of 54 which satisfy the criterion specified in the respective header.}\label{tab:physdispersion}
\end{table*}
Looking at the dispersion between OM I and OM II, 36 clusters are within one standard deviation, four within two and just four are more than three standard deviations away from each other. This implies that OM II seems to be in reasonable agreement with the two other models (usually in between the values from the other models). 



\subsection{Earth Mover's distance}
\label{subsec:emd}
The Earth Mover's distance (EMD), first introduced in \citet{rubner} is a "distance" function defined between two distributions. In the case where these distributions integrate over all space to the same value (e.g. they are probability distributions), the EMD is given in terms of the first Wasserstein distance \citep{levina}. 
A common analogy used to describe the EMD is the following: if the probability distributions are interpreted as two different ways of piling up a certain amount of earth, and the amount of earth at position $\vec{x}_{i}$ and $\vec{x}_{j}$ belonging to each probability distribution at those points are $Pr_{1}(\vec{x}_{i})$ and $Pr_{2}(\vec{x}_{j})$, then the EMD is the minimum cost of moving one pile into the other, where the cost of moving each "spadeful" is taken to be the mass of each spadeful ($f_{ij}$) $\times$ the distance by which it is moved ($|\vec{x_{i}} - \vec{x_{j}}|$).
For discrete two-dimensional probability distributions $Pr_{1}$ and $Pr_{2}$, with two-dimensional domains $\vec{x}_{i}$ and $\vec{y}_{j}$, then the EMD between these probability distributions $d_{\rm EMD}(Pr_{1}, Pr_{2})$ is defined to be the minimum value of
\begin{equation}\label{eqn:emd}
W(Pr_{1}, Pr_{2}) = \sum_{i = 1}^{m} \sum_{j = 1}^{n} f_{ij} |\vec{x_{i}} - \vec{y_{j}}|
\end{equation}
with respect to distance and $f_{ij}$.
Here $m$ and $n$ are the number of values in the domains of $Pr_{1}$ and $Pr_{2}$ respectively and $f_{ij}$ are the `flow' of probability density from $Pr_{1}(\vec{x}_{i})$ to $Pr_{2}(\vec{y}_{j})$. Different implementations of the algorithm use different distance measures, but we use the Euclidean distance in equation~\ref{eqn:emd}. The $f_{ij}$ are subject to the following constraints
\begin{equation}\label{eqn:emdconstraint1}
f_{ij} \geq 0, \, 1 \leq i \leq m, \, 1 \leq j \leq n;
\end{equation}
\begin{equation}\label{eqn:emdconstraint2}
\sum _{j=1}^{n}f_{ij} = Pr_{1}(\vec{x}_{i}), \, 1 \leq i \leq m;
\end{equation}
\begin{equation}\label{eqn:emdconstraint3}
\sum _{i=1}^{m}f_{ij} = Pr_{2}(\vec{y}_{j}), \, 1 \leq j \leq n;
\end{equation}
\begin{equation}\label{eqn:emdconstraint4}
\sum _{i = 1}^{m} \sum _{j = 1}^{n} f_{ij} = \sum _{i = 1}^{m} Pr_{1}(\vec{x}_{i}) = \sum _{j = 1}^{n}Pr_{2}(\vec{y}_{j}) = 1.
\end{equation}
For a more detailed account of the EMD see \citet{levina}.


\subsection{Application of EMD}
\label{subsec:EMD}

\begin{figure*}
  \begin{center}
  \includegraphics[ width=0.45\linewidth]{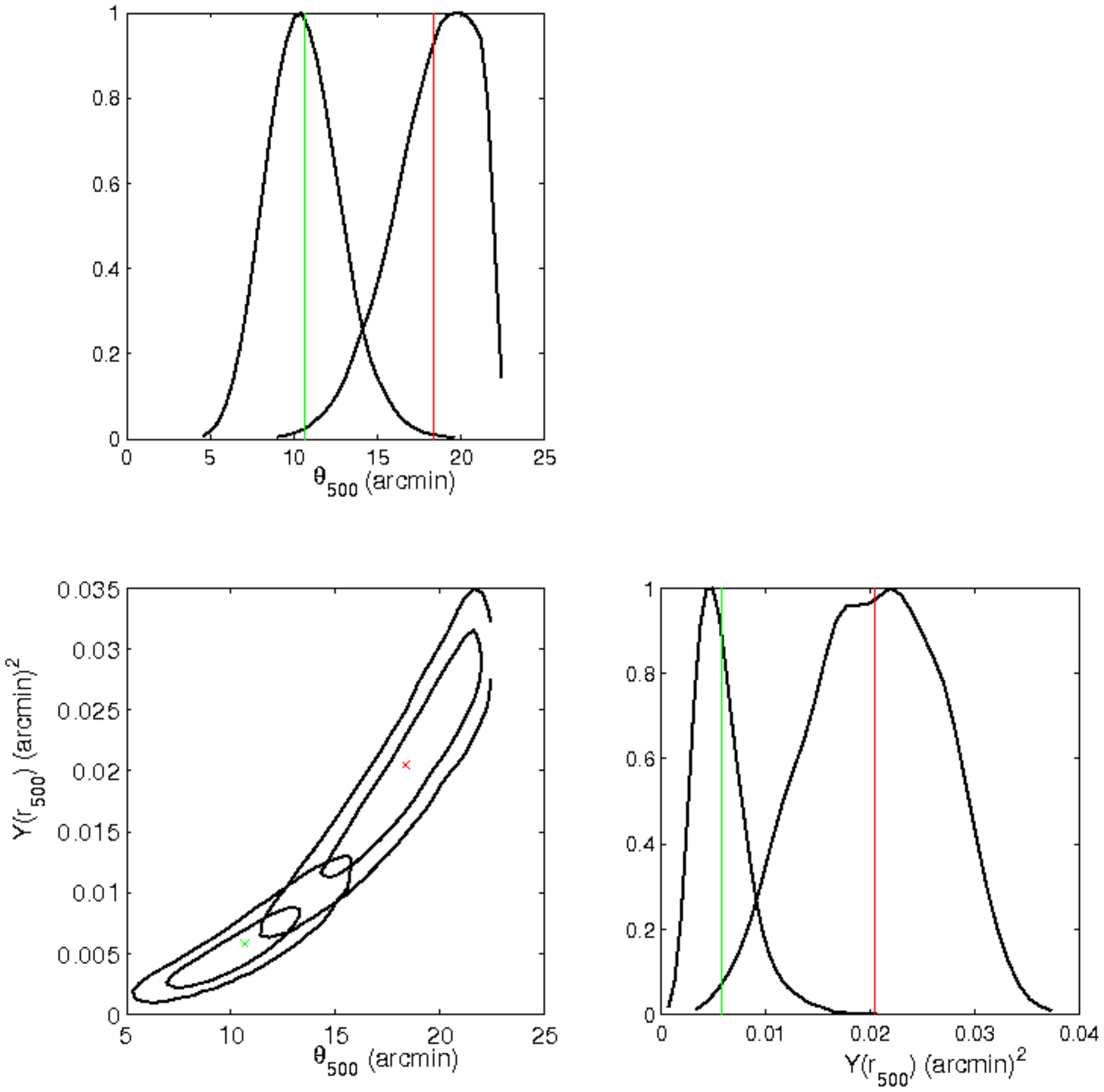}
  \includegraphics[ width=0.45\linewidth]{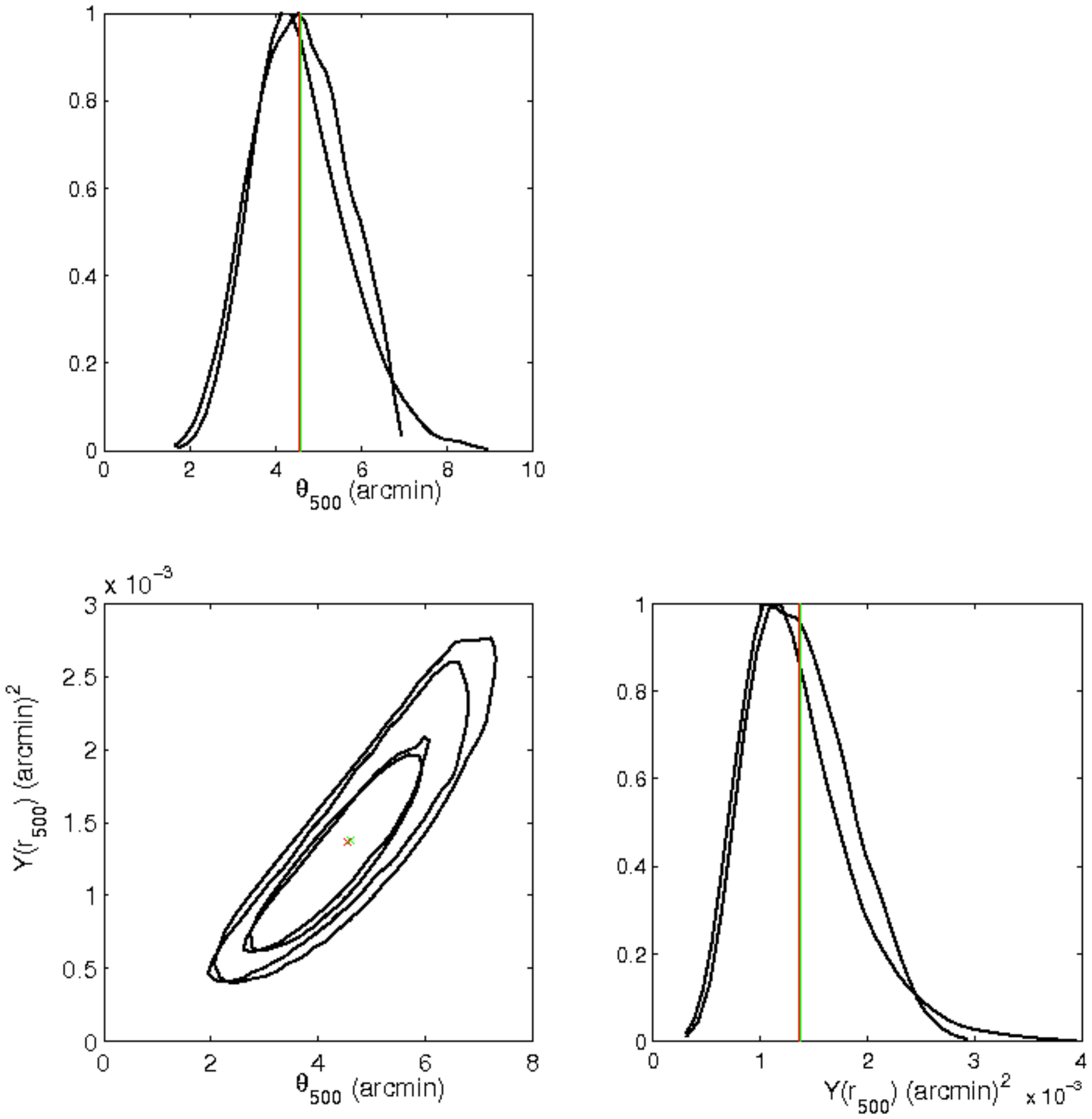}
  \medskip
  \centerline{(a) \hskip 0.45\linewidth (b)}
  \caption{(a) Highest $d_{\rm EMD}$ value $Y(r_{500})-\theta_{500}$ posteriors for cluster PSZ2~G044.20+48.66 at $z = 0.0894$. (b) Lowest $d_{\rm EMD}$ value $Y(r_{500})- \theta _{500}$ posteriors for cluster PSZ2~G132.47-17.27 at $z = 0.341$. For both triangle plots, the top graph shows the marginalised $\theta_{500}$ posteriors for OM II and OM I. The bottom right graph shows the marginalised $Y(r_{500})$ posteriors. The bottom left graph shows the two-dimensional $Y(r_{500})-\theta_{500}$ posteriors from which the EMD is calculated. The contours represent the 95\% and 68\% mean confidence intervals. Note that the parameters in the plots are not normalised, but the ones in the distance calculations are normalised by transforming the parameters as discussed in the text. For all of the plots, the green crosses / lines are the mean values of the OM I posteriors (the smaller values in (a)) and the red crosses / lines are the mean values of the OM II posteriors (the larger values in (a)). For Figure (b), the mean values for $Y(r_{500})$ are so close together that the lines cannot be distinguished.}
  \label{graph:highlow_emd}
  \end{center}
\end{figure*}

\begin{figure}
  \begin{center}
  \includegraphics[ width=0.90\linewidth]{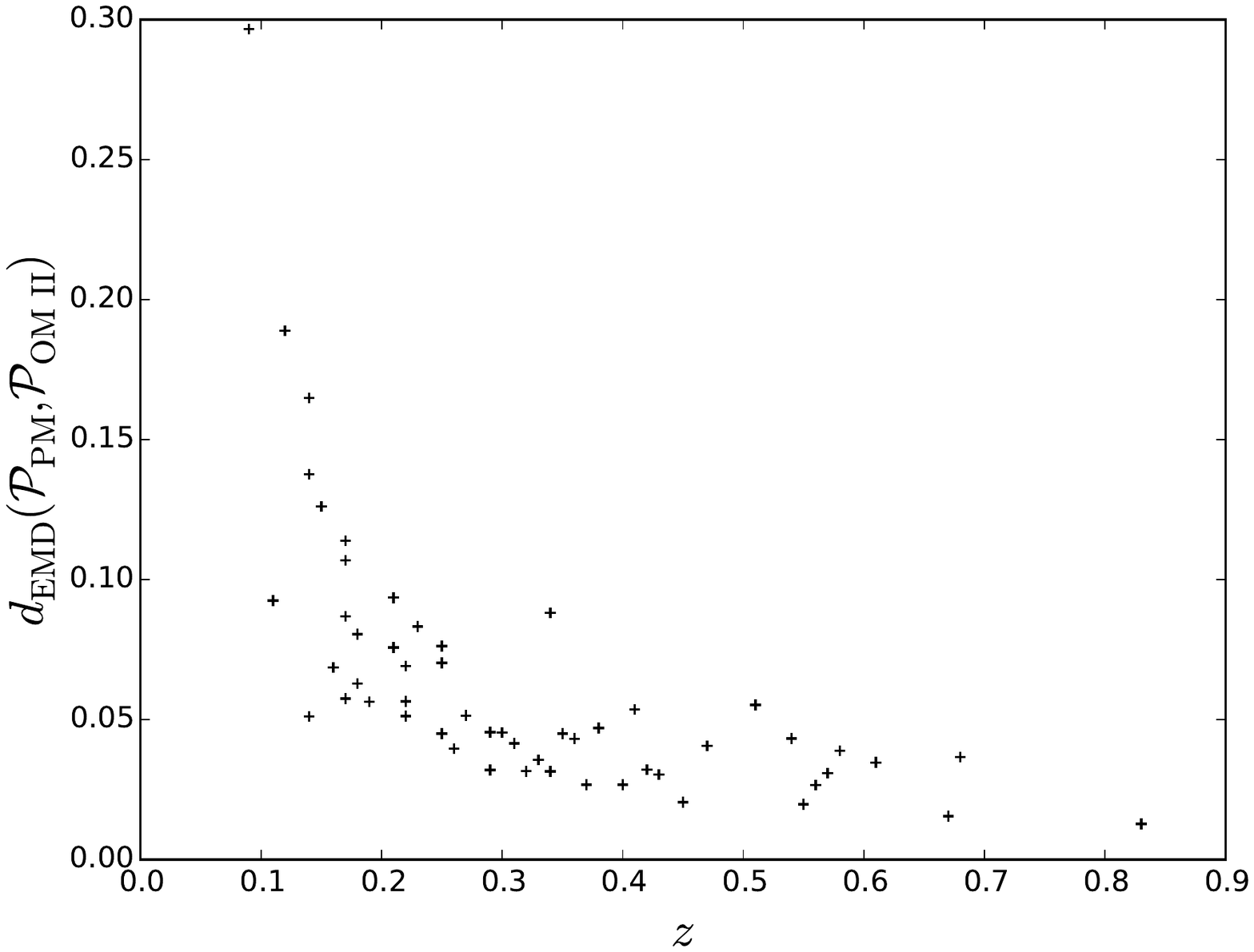}
  \caption{Earth Mover's distance calculated between $Y(r_{500})-\theta_{500}$ posteriors for PM and OM II, versus $z$ for the 54 clusters. The crosses indicate the point-- they are not error bars.}
\label{graph:physobsii_emd}
  \end{center}
\end{figure}

The EMD metric is applied to the different pairs of models using Gary Doran's wrapper\footnote{\url{https://github.com/garydoranjr/pyemd}.} for Yossi Rubner's algorithm \citep{rubner}. Before running the algorithm the $\left(Y(r_{500}), \, \theta_{500}\right)$ posteriors are normalised so that the metric is not skewed towards $\theta_{500}$ (the use of Euclidean distances in the EMD algorithm, are obviously misrepresentative if the dimensions are not normalised). Each dimension is normalised to the range $[0,1]$ by performing the following transformations
\begin{equation}\label{eqn:emdtransformations}
\theta_{500} \rightarrow \frac{\theta_{500} - \theta_{500, \, \rm min}}{\theta_{500, \, \rm max} - \theta_{500, \, \rm min}} ; Y(r_{500}) \rightarrow \frac{Y(r_{500}) - Y_{\rm min}(r_{500})}{Y_{\rm max}(r_{500}) - Y_{\rm min}(r_{500})}.
\end{equation}
The values for $\theta_{500, \, \rm min}$, $\theta_{500, \, \rm max}$, $Y_{\rm min}(r_{500})$ and $Y_{\rm max}(r_{500})$ are deduced by considering all of the values of $Y(r_{500})$ and $\theta_{\rm p}$ from the posteriors obtained from the three models at once, to ensure that all posterior values are normalised by the same factor. The larger the value of the EMD, the `further away' the distributions are from each other. The EMD was calculated for each cluster with each pair of models (giving $3 \times 54 = 162$ distances in total). The full set of EMD values calculated can be found in Table~\ref{tab:results2} in the Appendix. Table~\ref{tab:summary_emd} provides a summary of $d_{\rm EMD}(\mathcal{P}_{\rm PM}, \mathcal{P}_{\rm OM\, I})$, $d_{\rm EMD}(\mathcal{P}_{\rm OM\, I}, \mathcal{P}_{\rm OM\, II})$, $d_{\rm EMD}(\mathcal{P}_{\rm PM}, \mathcal{P}_{\rm OM\, II})$, and the union of the three. 
\begin{table*}
\centering
\begin{tabular}{{l}{c}{c}{c}{c}}
\hline
Statistic & $d_{\rm EMD}(\mathcal{P}_{\rm PM}, \mathcal{P}_{\rm OM\, I})$ & $d_{\rm EMD}(\mathcal{P}_{\rm PM}, \mathcal{P}_{\rm OM\, II})$ & $d_{\rm EMD}(\mathcal{P}_{\rm OM\, I}, \mathcal{P}_{\rm OM\, II})$ & union \\
\hline
mean & $ 0.093$ & $ 0.067$ & $ 0.057$ & $ 0.072$ \\
standard deviation & $ 0.057$ & $ 0.050$ & $ 0.077$ & $ 0.064$ \\
median & $ 0.076$ & $ 0.051$ & $ 0.027$ & $ 0.051$ \\
min & $ 0.020$ & $ 0.013$ & $ 0.006$ & $ 0.006$ \\
max & $ 0.225$ & $ 0.297$ & $ 0.514$ & $ 0.514$ \\
\hline
\end{tabular}
\caption{Summary of EMD values calculated between the $Y(r_{500})-\theta_{500}$ posterior distributions from all three model pairs, and their union.}\label{tab:summary_emd}
\end{table*}
Concerning both mean and median, the posteriors are most discrepant between the PM and OM I, followed by PM and OM II. However it is interesting to note that the two largest EMD values come from $d_{\rm EMD}(\mathcal{P}_{\rm OM\, II}, \mathcal{P}_{\rm OM\, I})$ and $d_{\rm EMD}(\mathcal{P}_{\rm PM}, \mathcal{P}_{\rm OM\, II})$ cases, with values $0.514$ and $0.297$ respectively. Furthermore these are from the same cluster, which is at the lowest $z$ ($= 0.0894$). This suggests that incorporating $z$ information into an observational model for very low redshift clusters has a significant effect. 
Ignoring the lowest redshift cluster (or by looking at the median value, which is skewed less by outliers), it is clear that of the three models, OM I and OM II posteriors are most in agreement with each other. Figure~\ref{graph:highlow_emd} shows the $Y(r_{500}), \, \theta_{500}$ posterior distributions created using \textsc{GetDist}\footnote{\url{http://getdist.readthedocs.io/en/latest/}.}, for the highest and lowest EMD values obtained from the 162 values calculated. Both of these come from OM II $-$ OM I comparisons. \\
Figure~\ref{graph:physobsii_emd} shows $d_{\rm EMD}(\mathcal{P}_{\rm PM}, \mathcal{P}_{\rm OM\, II})$ vs $z$ from which it is apparent that there is a negative correlation between $d_{\rm EMD}$ and $z$. 

\subsection{Physical and observational models comparison}

As described in Section~\ref{subsubsec:bayesmodel}, one can perform a model comparison, by comparing the Bayesian evidence values calculated when the models were applied to the same (AMI) datasets. We can also define the detection ratio of a model as the ratio of the evidences of the `data' and `null-data' runs. The first of these corresponds to modelling the cluster, background and detectable radio-sources. The null-data run models everything but the cluster. The ratio of these evidences therefore gives a measure of the significance that the cluster has in modelling the data. Note that the null-data run is the same for all three models considered here, as they only differ in the way they model the galaxy cluster itself. Table~\ref{tab:results2} in the Appendix gives the log of a detection ratio, $\ln (\mathcal{Z}_{i} / \mathcal{Z}_{\rm null})$ for each of the three models, and the ratios between the different pairs of models, $\ln (\mathcal{Z}_{i} / \mathcal{Z}_{j})$ where $\mathcal{Z}_{i}$ and $\mathcal{Z}_{j}$ are one of $\mathcal{Z}_{\rm{PM}}$, $\mathcal{Z}_{\rm{OM I}}$ or $\mathcal{Z}_{\rm{OM II}}$, for each cluster.


\subsubsection{Physical model and observational model I} 

\begin{figure*}
  \begin{center}
  \includegraphics[ width=0.45\linewidth]{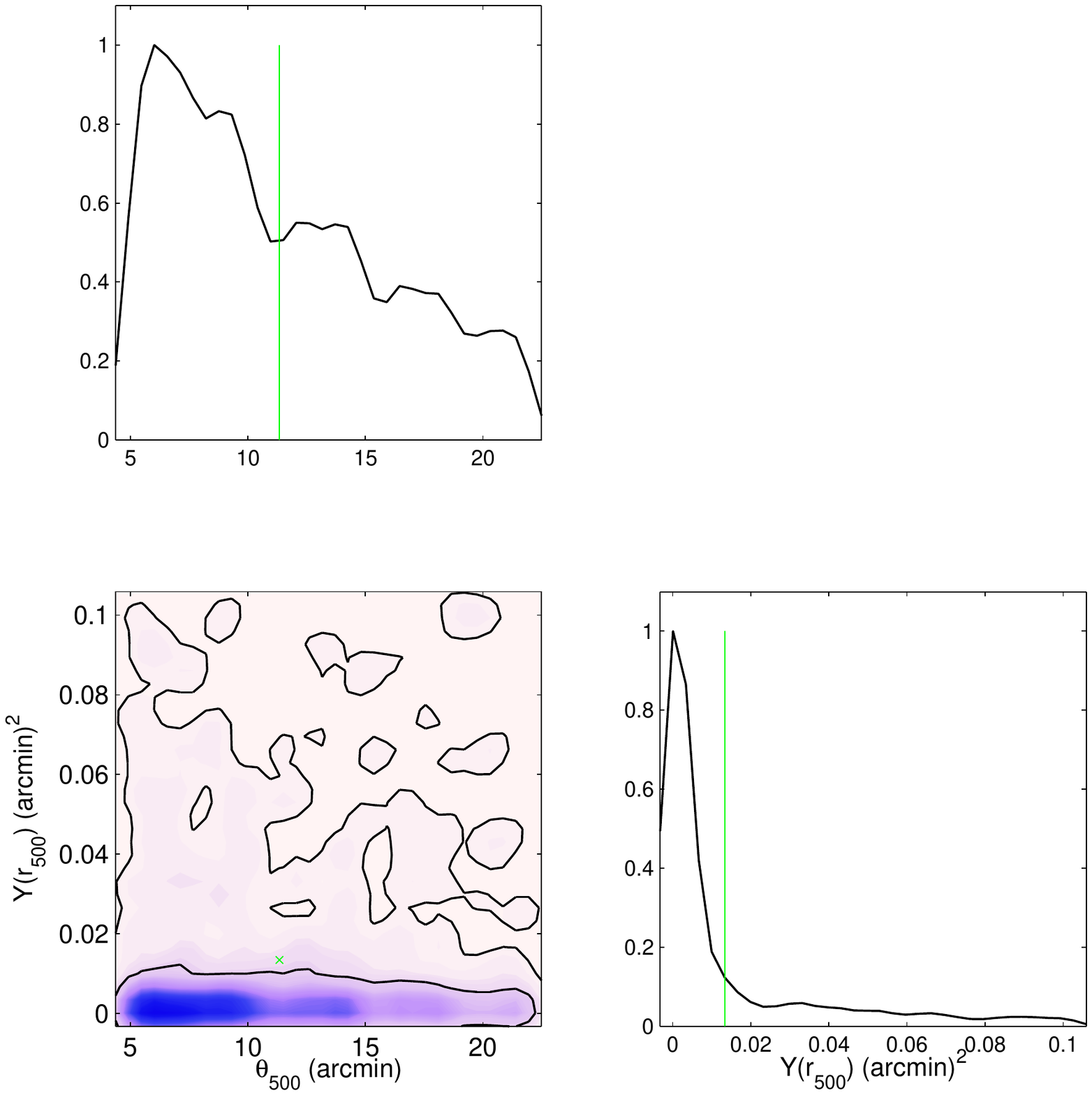}
  \includegraphics[ width=0.45\linewidth]{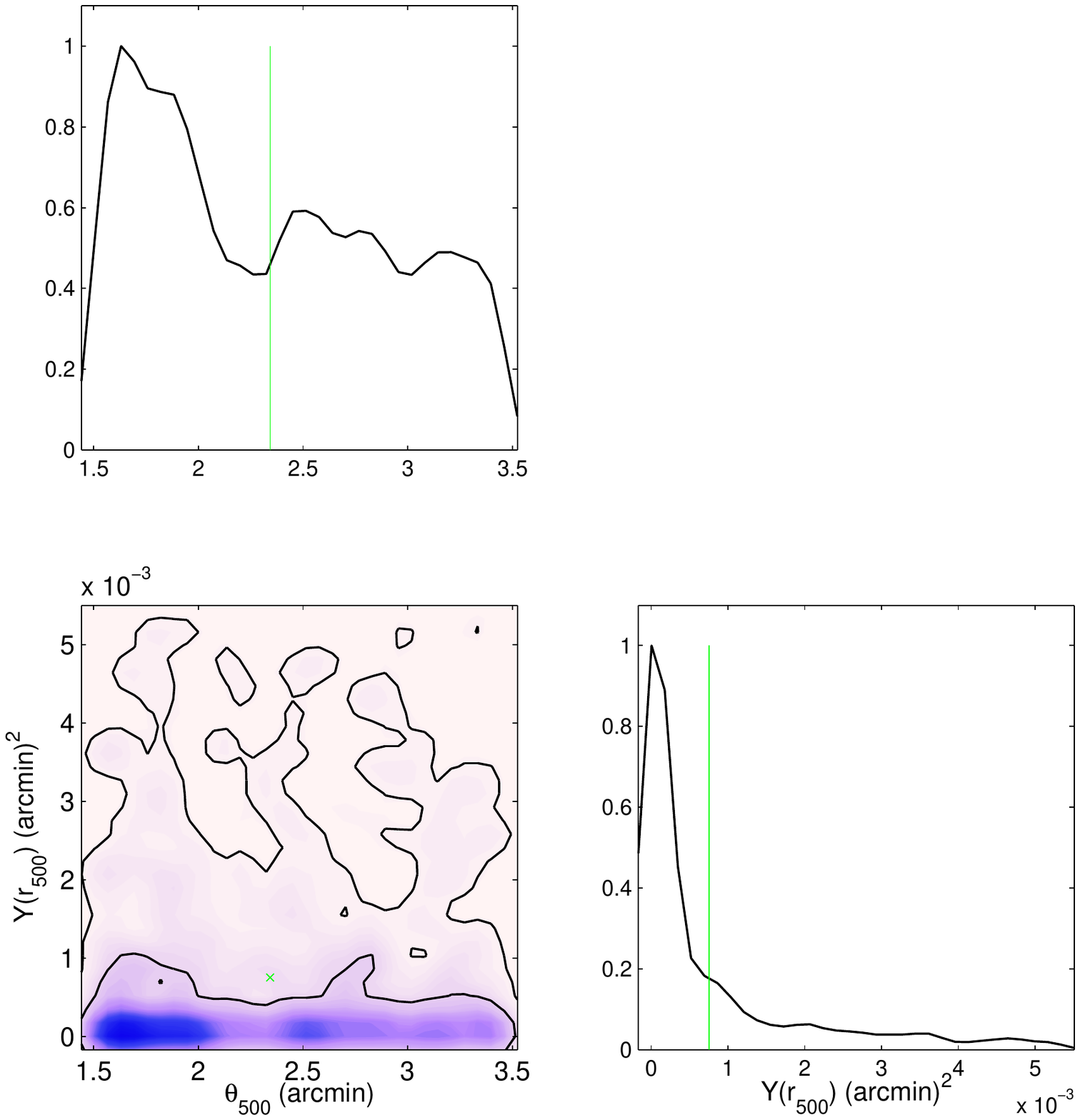}
  \medskip
  \centerline{(a) \hskip 0.45\linewidth (b)}
  \caption{(a) Lowest $z$ ($= 0.0894$) prior parameter space for $Y(r_{500})-\theta_{500}$ for the PM and OM II. (b) Highest $z$ ($= 0.83$) prior parameter space for $Y(r_{500})-\theta_{500}$ using the PM. Note the scales on the axes are different for each plot, and the green vertical lines correspond to the mean values.}
  \label{graph:phys_priors}
  \end{center}
\end{figure*}

The data favour OM I over the PM for 50 of the 54 clusters. Though in 36 of the 50 cases $\log(\mathcal{Z}_{\rm{PM}} / \mathcal{Z}_{\rm{OM I}})$ is between minus one and zero, which according to the Jeffreys scale means "more data are needed to come to a meaningful conclusion". (see Table~\ref{tab:jeffreys}). A further 12 of these had $\log(\mathcal{Z}_{\rm{PM}} / \mathcal{Z}_{\rm{OM I}})$ values between $-2.5$ and $-1$ which can be interpreted as "weak preference" in favour of OM I, whilst no clusters had a value of $\log(\mathcal{Z}_{\rm{PM}} / \mathcal{Z}_{\rm{OM I}})$ less than minus five ("strong preference" in favour of OM I). The largest absolute value for the ratio was actually in favour of the PM with $\ln (\mathcal{Z}_{\rm PM} / \mathcal{Z}_{\rm OM\, I}) = 4.73 \pm 0.23$ (for the lowest $z$ cluster) which suggests "moderate preference" towards the PM. There is no correlation between $\log(\mathcal{Z}_{\rm{PM}} / \mathcal{Z}_{\rm{OM I}})$ and $z$. \\ Figure~\ref{graph:phys_priors} shows the prior space for the observational parameters corresponding to the PM with the lowest and highest $z$ values in the sample.

\subsubsection{Observational models I and II}

\begin{figure}
  \begin{center}
  \includegraphics[ width=0.90\linewidth]{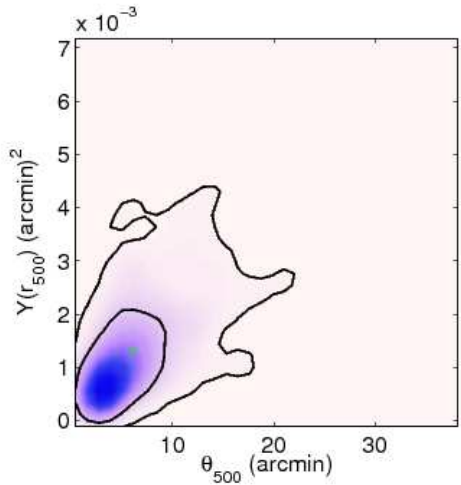} 
  \caption{Two-dimensional prior probability distribution of $Y(r_{500})$ and $\theta_{500}$ for OM I, which is based on \textit{Planck} data as detailed in Section~\ref{subsec:obs_i}.}
\label{graph:obsiobsii_priors}
  \end{center}
\end{figure}

Similarly, OM I is favoured over OM II for 53 clusters, but with 14 cases having  $0 \leq \log(\mathcal{Z}_{\rm{OM I}} / \mathcal{Z}_{\rm{OM II}}) \leq 1$. Again the highest absolute value came from the lowest redshift cluster, highlighting the importance of $z$ information at such a low $z$ value.
Since these models have the same input parameters, it is easier to compare their sampling parameter spaces.
Figure~\ref{graph:obsiobsii_priors} shows the prior range of $\left(Y(r_{500}), \, \theta_{500}\right)$ for OM I. Around 68\% of the prior mass (i.e. the inner contour in the Figure) is bounded roughly by $Y(r_{500}) = 2\times 10^{-3}~\rm{arcmin}^{2}$ and $\theta_{500} = 10~\rm{arcmin}$. The 95\% contour gives upper bounds of $Y(r_{500}) \approx 4\times10^{-3}~\rm{arcmin}^{2}$ and $\theta_{500} \approx 15~\rm{arcmin}$. In comparison the OM II prior ranges for the lowest redshift cluster are $\theta_{500} = [4.9,~19.0]~\rm{arcmin}$ and $Y(r_{500}) = [0.006,~1.0]\times 10^{-1}~\rm{arcmin}^{2}$, and for the highest redshift cluster are $ \theta_{500} = [0.8,~3.5]~\rm{arcmin}$, $ Y(r_{500}) = [0.003,~5.0]\times 10^{-3}~\rm{arcmin}^{2}$. The ratio of the upper and lower limits for $\theta$ and $Y$ are approximately $4.5$ and $1.8 \times 10^{3}$ across all clusters. This suggests that the ratio of the bounds of the parameter space for each cluster does not change for the OM II, but that the sampling space is shifted depending on $z$.
Note that even though the sampling parameters for the observational models are $Y_{\rm tot}$ and $\theta_{\rm p}$, these are related to $Y(r_{500})$ and $\theta_{500}$ by constant factors, and so comparisons made on both are equivalent.

\subsubsection{Physical model and observational model II}

Comparison of PM and OM II, the models which incorporate redshift information into their priors leads to interesting results. For 43 clusters, the PM is preferred over OM II. However for all of these clusters $\log(\mathcal{Z}_{\rm{PM}} / \mathcal{Z}_{\rm{OM II}})$ is less than one, meaning that none of them give "conclusive" model preference. There are only three clusters which give "weak evidence" in favour of a model (OM II). These are the clusters at redshift $z = 0.144, \, 0.341, \, 0.5131$ with ratio values $ -1.88, \, -1.06, \, -1.16$ respectively. The fact that data from 51 clusters do not provide any "conclusive" preference between PM and OM II, suggests that these models are equally well suited for the current data, even though their parameter estimates are often not in such agreement.


\section{Conclusions}
\label{sec:summary}

For the cluster sample analysed in \citet{2019MNRAS.483.3529J} (KJ19), we compare the parameter estimates obtained from different physical and observational models applied to AMI data using Bayesian analysis. 
The physical model (PM) used is as described in KJ19, and the observational models (OM I and OM II) are based on the one described in \citet{2015A&A...580A..95P}. 
We have focused on comparisons of $Y(r_{500})$, and found the following.
\begin{itemize}
\item The PM generally yields lower estimates of $Y$ relative to the observational models, apart from at low $z$ where the reverse is true. 
\item For two thirds of the sample, the OM I and OM II estimates are within one combined standard deviation of each other.
\end{itemize}
To investigate further the discrepancies between the three models, we computed the Earth Mover's distance between the two-dimensional posterior distributions in $Y(r_{500}), \, \theta_{\rm 500}$ space, for each model pair. This gives a measure of the `distance' between the respective probability distributions. We then compared the evidence values obtained from the Bayesian analysis of the AMI data using the different models, referring to the Jeffreys scale to form conclusions on model preference. We found the following.
\begin{itemize}
\item Based on the Earth Mover's distances calculated for each cluster, the posteriors are most discrepant between the PM and OM I models when the sample was considered as a whole, followed by PM and OM II.
\item The two largest discrepancies come from the lowest-$z$ cluster, one between PM and OM I and one between OM II and OM I, suggesting that $z$ information at very low $z$ can have a large effect on the different models.
\item The distance between posteriors from PM and OM II clearly decreases with increasing $z$. This suggests that the difference between physical and observational model parameter estimates, provided the latter also includes $z$ information, is reduced at higher $z$.
\item When comparing Bayesian evidence values, OM I is preferred over PM for 50 of the clusters, although only 14 of these showed either "weak" or "moderate" preference to OM I (the remaining 36 being "inconclusive"); however the highest $\log(\mathrm{evidence} \, \mathrm{ratio})$ actually favours the PM ("moderate" preference) and occurs for the lowest-$z$ cluster.
\item Similarly, OM I is preferred to OM II in 53 of the cases. 14 suggested more data are needed to come to a "meaningful" conclusion, while the remaining 39 clusters showed "weak" or "moderate" preference for OM I. This suggests that OM I is the preferred model in more cases relative to OM II than when OM I is compared with PM.
\item For 43 of the clusters, PM is preferred over OM II; however in all of these cases, the Jeffreys scale suggests "no conclusion can be made without more data", and only three clusters give any "conclusive" preference (a "weak" preference in favour for OM II).
\end{itemize}
In forthcoming papers \citep{2019MNRAS.489.3135J} and \citep{2019arXiv190900029J}, different dark matter and pressure models will be considered, and in \citet{2019MNRAS.486.2116P}, Bayesian analysis will be performed on joint AMI-\textit{Planck} datasets.


\section*{Acknowledgements}
This work was performed using the Darwin Supercomputer of the University of Cambridge High Performance Computing (HPC) Service (\url{http://www.hpc.cam.ac.uk/}), provided by Dell Inc. using Strategic Research Infrastructure Funding from the Higher Education Funding Council for England and funding from the Science and Technology Facilities Council. The authors would like to thank Stuart Rankin from HPC and Greg Willatt and David Titterington from Cavendish Astrophysics for computing assistance. They would also like to thank Dave Green for his invaluable help using \LaTeX.
Kamran Javid acknowledges an STFC studentship. Yvette Perrott acknowledges support from a Trinity College Junior Research Fellowship.


\setlength{\bibsep}{0pt}            
\renewcommand{\bibname}{References} 


\newgeometry{margin=1cm}
\onecolumn
\begin{landscape}
\appendix
\section{Results tables}\label{sec:results_tables}
\begin{center}
\begin{longtable}{llllllllll}
\caption{Summary of parameter estimates for final sample of 54 clusters. 
All $Y$ values are given in units of $\times 10^{-3}~\rm (arcmin^{2})$, and all cluster centre coordinates are given in arcseconds. The cluster centre estimates from the physical model are omitted here but can be found in the results Table in Appendix~A of KJ19, which is ordered in the same order as this Table. Note the Table in KJ19 also gives external names associated with these clusters, as well as the method used to measure the respective redshifts (i.e. spectroscopic or photometric). 
}\label{tab:results1} \\

\hline \multicolumn{1}{c}{Row} & \multicolumn{1}{c}{\textit{Planck} ID}  & \multicolumn{1}{c}{$z$} & \multicolumn{1}{c}{$Y_{\rm PM}(r_{500})$} & \multicolumn{1}{c}{$Y_{\rm OM\, I}(r_{500})$} & \multicolumn{1}{c}{$x_{0, \rm OM I}$} & \multicolumn{1}{c}{$y_{0, \rm OM\, I}$} & \multicolumn{1}{c}{$Y_{\rm OM\, II}(r_{500})$} & \multicolumn{1}{c}{$x_{0, \rm OM\, II}$} & \multicolumn{1}{c}{$y_{0, \rm OM\, II}$} \\ \hline 
\endfirsthead

\multicolumn{10}{c}%
{{\tablename\ \thetable{} -- continued from previous page}} \\
\hline \multicolumn{1}{c}{Row} & \multicolumn{1}{c}{\textit{Planck} ID}  & \multicolumn{1}{c}{$z$} & \multicolumn{1}{c}{$Y_{\rm PM}(r_{500})$} & \multicolumn{1}{c}{$Y_{\rm OM\, I}(r_{500})$} & \multicolumn{1}{c}{$x_{0, \rm OM I}$} & \multicolumn{1}{c}{$y_{0, \rm OM\, I}$} & \multicolumn{1}{c}{$Y_{\rm OM\, II}(r_{500})$} & \multicolumn{1}{c}{$x_{0, \rm OM\, II}$} & \multicolumn{1}{c}{$y_{0, \rm OM\, II}$} \\ \hline 
\endhead

\tabulinesep=_1mm
\extrarowsep=1mm
\LTcapwidth=\textwidth

1	&	PSZ2~G044.20+48.66	&	$	0.0894	$	&	$	11.59	\pm	2.28	$	&	$	6.77	\pm	3.32	$	&	$	6.53	\pm	18.56	$	&	$	8.93	\pm	14.41	$	&	$	20.48	\pm	6.19	$	&	$	10.36	\pm	18.38	$	&	$	8.32	\pm	15.32	$	\\
2	&	PSZ2~G053.53+59.52	&	$	0.113	$	&	$	3.81	\pm	0.67	$	&	$	2.02	\pm	0.90	$	&	$	-1.77	\pm	12.69	$	&	$	23.19	\pm	9.38	$	&	$	3.12	\pm	1.74	$	&	$	-1.07	\pm	12.67	$	&	$	20.89	\pm	9.88	$	\\
3	&	PSZ2~G151.90+11.63	&	$	0.12	$	&	$	1.76	\pm	0.50	$	&	$	2.55	\pm	1.56	$	&	$	63.93	\pm	28.11	$	&	$	67.61	\pm	18.86	$	&	$	4.09	\pm	1.83	$	&	$	59.05	\pm	27.67	$	&	$	67.19	\pm	19.36	$	\\
4	&	PSZ2~G218.59+71.31	&	$	0.137	$	&	$	0.45	\pm	0.25	$	&	$	0.35	\pm	0.15	$	&	$	-8.85	\pm	14.58	$	&	$	-17.72	\pm	14.59	$	&	$	0.43	\pm	0.27	$	&	$	0.04	\pm	23.62	$	&	$	-16.95	\pm	24.66	$	\\
5	&	PSZ2~G226.18+76.79	&	$	0.1427	$	&	$	2.45	\pm	0.45	$	&	$	0.91	\pm	0.45	$	&	$	-45.20	\pm	10.61	$	&	$	6.46	\pm	12.25	$	&	$	1.21	\pm	0.51	$	&	$	-42.92	\pm	10.66	$	&	$	3.80	\pm	12.00	$	\\
6	&	PSZ2~G165.06+54.13	&	$	0.144	$	&	$	2.26	\pm	0.54	$	&	$	0.70	\pm	0.25	$	&	$	29.82	\pm	10.17	$	&	$	-29.36	\pm	12.22	$	&	$	0.95	\pm	0.27	$	&	$	31.51	\pm	10.76	$	&	$	-29.04	\pm	12.83	$	\\
7	&	PSZ2~G077.90-26.63	&	$	0.147	$	&	$	2.80	\pm	0.46	$	&	$	1.35	\pm	0.48	$	&	$	-27.99	\pm	9.91	$	&	$	20.12	\pm	11.23	$	&	$	1.48	\pm	0.49	$	&	$	-28.06	\pm	10.13	$	&	$	19.93	\pm	11.07	$	\\
8	&	PSZ2~G050.40+31.17	&	$	0.164	$	&	$	1.01	\pm	0.29	$	&	$	1.07	\pm	0.70	$	&	$	37.21	\pm	20.82	$	&	$	9.59	\pm	19.09	$	&	$	1.18	\pm	0.76	$	&	$	36.11	\pm	22.25	$	&	$	9.30	\pm	19.70	$	\\
9	&	PSZ2~G097.72+38.12	&	$	0.1709	$	&	$	2.65	\pm	0.46	$	&	$	2.72	\pm	1.26	$	&	$	29.79	\pm	15.43	$	&	$	-2.59	\pm	13.68	$	&	$	3.97	\pm	1.49	$	&	$	32.13	\pm	15.62	$	&	$	-1.56	\pm	13.81	$	\\
10	&	PSZ2~G099.30+20.92	&	$	0.171	$	&	$	0.97	\pm	0.31	$	&	$	0.79	\pm	0.49	$	&	$	-35.09	\pm	19.11	$	&	$	-24.57	\pm	21.53	$	&	$	0.86	\pm	0.51	$	&	$	-36.16	\pm	19.13	$	&	$	-25.55	\pm	21.67	$	\\
11	&	PSZ2~G067.17+67.46	&	$	0.1712	$	&	$	2.70	\pm	0.46	$	&	$	1.30	\pm	0.54	$	&	$	34.00	\pm	11.65	$	&	$	-30.54	\pm	10.97	$	&	$	1.48	\pm	0.60	$	&	$	33.18	\pm	11.61	$	&	$	-31.32	\pm	11.16	$	\\
12	&	PSZ2~G167.67+17.63	&	$	0.174	$	&	$	0.72	\pm	0.30	$	&	$	1.69	\pm	1.05	$	&	$	-24.86	\pm	32.03	$	&	$	10.55	\pm	28.11	$	&	$	1.33	\pm	0.77	$	&	$	-23.41	\pm	33.17	$	&	$	11.93	\pm	29.04	$	\\
13	&	PSZ2~G066.68+68.44	&	$	0.181	$	&	$	0.72	\pm	0.29	$	&	$	1.24	\pm	0.79	$	&	$	55.97	\pm	25.19	$	&	$	9.20	\pm	32.13	$	&	$	1.12	\pm	0.72	$	&	$	56.41	\pm	26.70	$	&	$	7.31	\pm	32.63	$	\\
14	&	PSZ2~G065.28+44.53	&	$	0.183	$	&	$	0.79	\pm	0.28	$	&	$	0.65	\pm	0.38	$	&	$	-21.13	\pm	20.72	$	&	$	-15.63	\pm	18.96	$	&	$	0.61	\pm	0.34	$	&	$	-19.57	\pm	22.13	$	&	$	-16.08	\pm	20.64	$	\\
15	&	PSZ2~G084.47+12.63	&	$	0.185	$	&	$	0.67	\pm	0.25	$	&	$	0.58	\pm	0.29	$	&	$	-67.12	\pm	29.59	$	&	$	-23.26	\pm	18.01	$	&	$	0.53	\pm	0.28	$	&	$	-69.03	\pm	30.83	$	&	$	-20.78	\pm	20.07	$	\\
16	&	PSZ2~G100.04+23.73	&	$	0.21	$	&	$	0.65	\pm	0.18	$	&	$	1.28	\pm	0.75	$	&	$	17.47	\pm	19.11	$	&	$	-22.73	\pm	22.21	$	&	$	1.05	\pm	0.55	$	&	$	17.93	\pm	19.85	$	&	$	-23.27	\pm	22.53	$	\\
17	&	PSZ2~G180.60+76.65	&	$	0.2138	$	&	$	0.63	\pm	0.20	$	&	$	1.73	\pm	0.93	$	&	$	36.57	\pm	16.66	$	&	$	-73.38	\pm	20.39	$	&	$	1.11	\pm	0.50	$	&	$	35.90	\pm	17.29	$	&	$	-70.57	\pm	22.18	$	\\
18	&	PSZ2~G166.09+43.38	&	$	0.2172	$	&	$	1.67	\pm	0.28	$	&	$	1.10	\pm	0.50	$	&	$	-4.29	\pm	10.57	$	&	$	-6.54	\pm	9.54	$	&	$	1.14	\pm	0.46	$	&	$	-4.73	\pm	10.32	$	&	$	-6.66	\pm	9.63	$	\\
19	&	PSZ2~G125.30-27.99	&	$	0.223	$	&	$	0.45	\pm	0.18	$	&	$	0.99	\pm	0.64	$	&	$	-8.12	\pm	26.53	$	&	$	2.49	\pm	30.79	$	&	$	0.60	\pm	0.38	$	&	$	-9.03	\pm	28.36	$	&	$	6.48	\pm	31.71	$	\\
20	&	PSZ2~G060.13+11.44	&	$	0.224	$	&	$	1.00	\pm	0.20	$	&	$	1.17	\pm	0.64	$	&	$	-64.93	\pm	12.76	$	&	$	-49.60	\pm	15.02	$	&	$	1.12	\pm	0.56	$	&	$	-64.67	\pm	12.69	$	&	$	-49.56	\pm	14.77	$	\\
21	&	PSZ2~G166.62+42.13	&	$	0.232	$	&	$	0.29	\pm	0.13	$	&	$	1.57	\pm	0.92	$	&	$	-36.13	\pm	30.51	$	&	$	-54.22	\pm	32.52	$	&	$	0.53	\pm	0.35	$	&	$	-34.92	\pm	31.92	$	&	$	-40.79	\pm	38.30	$	\\
22	&	PSZ2~G097.94+19.43	&	$	0.25	$	&	$	0.45	\pm	0.17	$	&	$	1.24	\pm	0.69	$	&	$	-121.19	\pm	21.52	$	&	$	-2.42	\pm	32.74	$	&	$	0.73	\pm	0.41	$	&	$	-115.20	\pm	27.60	$	&	$	-5.84	\pm	34.15	$	\\
23	&	PSZ2~G164.29+08.94	&	$	0.251	$	&	$	0.59	\pm	0.13	$	&	$	0.87	\pm	0.43	$	&	$	-62.15	\pm	13.92	$	&	$	20.46	\pm	17.35	$	&	$	0.73	\pm	0.35	$	&	$	-62.23	\pm	13.90	$	&	$	18.67	\pm	17.99	$	\\
24	&	PSZ2~G133.60+69.04	&	$	0.254	$	&	$	0.47	\pm	0.20	$	&	$	1.60	\pm	1.12	$	&	$	0.13	\pm	24.80	$	&	$	66.74	\pm	35.89	$	&	$	0.80	\pm	0.45	$	&	$	3.35	\pm	25.98	$	&	$	63.00	\pm	37.13	$	\\
25	&	PSZ2~G086.47+15.31	&	$	0.26	$	&	$	1.48	\pm	0.33	$	&	$	1.70	\pm	0.71	$	&	$	-41.40	\pm	14.66	$	&	$	19.45	\pm	13.73	$	&	$	1.58	\pm	0.60	$	&	$	-40.08	\pm	14.39	$	&	$	20.08	\pm	13.75	$	\\
26	&	PSZ2~G139.62+24.18	&	$	0.2671	$	&	$	0.89	\pm	0.16	$	&	$	0.77	\pm	0.34	$	&	$	35.74	\pm	11.80	$	&	$	-13.45	\pm	11.11	$	&	$	0.70	\pm	0.33	$	&	$	35.83	\pm	11.49	$	&	$	-13.78	\pm	10.76	$	\\
27	&	PSZ2~G184.68+28.91	&	$	0.288	$	&	$	0.76	\pm	0.12	$	&	$	0.95	\pm	0.38	$	&	$	22.66	\pm	10.55	$	&	$	12.19	\pm	10.37	$	&	$	0.83	\pm	0.31	$	&	$	22.58	\pm	10.48	$	&	$	13.03	\pm	10.41	$	\\
28	&	PSZ2~G154.13+40.19	&	$	0.29	$	&	$	0.55	\pm	0.13	$	&	$	0.72	\pm	0.50	$	&	$	71.59	\pm	15.07	$	&	$	-42.78	\pm	13.41	$	&	$	0.46	\pm	0.23	$	&	$	69.88	\pm	14.52	$	&	$	-42.45	\pm	13.20	$	\\
29	&	PSZ2~G095.49+16.41	&	$	0.3	$	&	$	0.39	\pm	0.12	$	&	$	0.87	\pm	0.54	$	&	$	-19.80	\pm	21.12	$	&	$	-94.58	\pm	19.43	$	&	$	0.48	\pm	0.26	$	&	$	-22.58	\pm	20.72	$	&	$	-98.75	\pm	20.62	$	\\
30	&	PSZ2~G109.52-19.16	&	$	0.3092	$	&	$	0.78	\pm	0.16	$	&	$	1.00	\pm	0.57	$	&	$	-31.66	\pm	14.34	$	&	$	-15.21	\pm	15.68	$	&	$	0.82	\pm	0.39	$	&	$	-31.16	\pm	14.43	$	&	$	-15.23	\pm	15.95	$	\\
31	&	PSZ2~G198.90+18.16	&	$	0.3184	$	&	$	0.62	\pm	0.12	$	&	$	0.86	\pm	0.40	$	&	$	27.42	\pm	15.36	$	&	$	-59.55	\pm	12.35	$	&	$	0.69	\pm	0.27	$	&	$	27.03	\pm	15.25	$	&	$	-57.65	\pm	12.64	$	\\
32	&	PSZ2~G152.33+81.28	&	$	0.333	$	&	$	0.43	\pm	0.11	$	&	$	0.78	\pm	0.42	$	&	$	-49.96	\pm	20.35	$	&	$	44.73	\pm	15.45	$	&	$	0.48	\pm	0.22	$	&	$	-53.60	\pm	20.93	$	&	$	43.79	\pm	15.28	$	\\
33	&	PSZ2~G108.17-11.56	&	$	0.336	$	&	$	0.61	\pm	0.12	$	&	$	2.24	\pm	1.10	$	&	$	27.48	\pm	14.92	$	&	$	-36.56	\pm	20.44	$	&	$	1.12	\pm	0.25	$	&	$	30.62	\pm	13.89	$	&	$	-51.07	\pm	19.77	$	\\
34	&	PSZ2~G132.47-17.27	&	$	0.341	$	&	$	1.25	\pm	0.21	$	&	$	1.38	\pm	0.52	$	&	$	32.53	\pm	10.83	$	&	$	16.82	\pm	12.65	$	&	$	1.37	\pm	0.47	$	&	$	32.34	\pm	10.66	$	&	$	16.61	\pm	12.56	$	\\
35	&	PSZ2~G207.88+81.31	&	$	0.353	$	&	$	1.05	\pm	0.18	$	&	$	0.90	\pm	0.34	$	&	$	67.45	\pm	8.46	$	&	$	61.30	\pm	11.45	$	&	$	0.82	\pm	0.29	$	&	$	66.90	\pm	8.21	$	&	$	59.84	\pm	11.43	$	\\
36	&	PSZ2~G157.32-26.77	&	$	0.356	$	&	$	1.52	\pm	0.27	$	&	$	1.25	\pm	0.42	$	&	$	-0.28	\pm	8.01	$	&	$	19.15	\pm	11.86	$	&	$	1.23	\pm	0.39	$	&	$	-1.07	\pm	7.59	$	&	$	17.73	\pm	11.58	$	\\
37	&	PSZ2~G071.21+28.86	&	$	0.366	$	&	$	0.72	\pm	0.15	$	&	$	0.91	\pm	0.34	$	&	$	-29.47	\pm	10.86	$	&	$	-12.29	\pm	14.04	$	&	$	0.75	\pm	0.25	$	&	$	-29.64	\pm	10.48	$	&	$	-12.13	\pm	13.74	$	\\
38	&	PSZ2~G194.98+54.12	&	$	0.375	$	&	$	0.65	\pm	0.15	$	&	$	1.28	\pm	0.69	$	&	$	32.85	\pm	12.59	$	&	$	-5.89	\pm	18.85	$	&	$	0.93	\pm	0.32	$	&	$	32.71	\pm	12.45	$	&	$	-3.46	\pm	19.90	$	\\
39	&	PSZ2~G109.86+27.94	&	$	0.4	$	&	$	0.21	\pm	0.09	$	&	$	0.30	\pm	0.11	$	&	$	8.03	\pm	16.29	$	&	$	-1.95	\pm	14.87	$	&	$	0.20	\pm	0.07	$	&	$	7.15	\pm	21.69	$	&	$	2.87	\pm	17.97	$	\\
40	&	PSZ2~G083.29-31.03	&	$	0.412	$	&	$	0.95	\pm	0.17	$	&	$	0.66	\pm	0.21	$	&	$	75.26	\pm	13.22	$	&	$	-0.29	\pm	12.25	$	&	$	0.60	\pm	0.20	$	&	$	72.16	\pm	13.03	$	&	$	2.13	\pm	11.88	$	\\
41	&	PSZ2~G063.38+53.44	&	$	0.422	$	&	$	0.93	\pm	0.19	$	&	$	1.28	\pm	0.45	$	&	$	39.37	\pm	14.20	$	&	$	49.33	\pm	10.77	$	&	$	1.12	\pm	0.29	$	&	$	41.65	\pm	13.30	$	&	$	48.43	\pm	10.17	$	\\
42	&	PSZ2~G063.80+11.42	&	$	0.426	$	&	$	0.24	\pm	0.08	$	&	$	0.80	\pm	0.46	$	&	$	-42.04	\pm	23.06	$	&	$	-44.32	\pm	20.40	$	&	$	0.29	\pm	0.14	$	&	$	-36.98	\pm	23.28	$	&	$	-45.28	\pm	20.74	$	\\
43	&	PSZ2~G157.43+30.34	&	$	0.45	$	&	$	0.82	\pm	0.13	$	&	$	0.91	\pm	0.26	$	&	$	-61.41	\pm	7.56	$	&	$	4.85	\pm	8.34	$	&	$	0.85	\pm	0.23	$	&	$	-61.63	\pm	7.29	$	&	$	4.79	\pm	8.26	$	\\
44	&	PSZ2~G150.56+58.32	&	$	0.47	$	&	$	0.93	\pm	0.25	$	&	$	0.86	\pm	0.38	$	&	$	9.81	\pm	14.03	$	&	$	35.97	\pm	18.29	$	&	$	0.70	\pm	0.25	$	&	$	8.34	\pm	12.93	$	&	$	36.51	\pm	18.01	$	\\
45	&	PSZ2~G170.98+39.45	&	$	0.5131	$	&	$	0.54	\pm	0.08	$	&	$	1.62	\pm	0.68	$	&	$	23.91	\pm	12.09	$	&	$	-18.32	\pm	13.31	$	&	$	0.88	\pm	0.17	$	&	$	26.68	\pm	11.52	$	&	$	-22.95	\pm	12.68	$	\\
46	&	PSZ2~G094.56+51.03	&	$	0.5392	$	&	$	0.63	\pm	0.10	$	&	$	0.50	\pm	0.09	$	&	$	82.24	\pm	7.64	$	&	$	50.61	\pm	8.76	$	&	$	0.45	\pm	0.08	$	&	$	81.87	\pm	7.67	$	&	$	50.51	\pm	8.62	$	\\
47	&	PSZ2~G228.16+75.20	&	$	0.545	$	&	$	1.06	\pm	0.10	$	&	$	1.35	\pm	0.27	$	&	$	-14.53	\pm	5.57	$	&	$	16.35	\pm	5.31	$	&	$	1.25	\pm	0.21	$	&	$	-14.39	\pm	5.59	$	&	$	16.50	\pm	5.08	$	\\
48	&	PSZ2~G213.39+80.59	&	$	0.5586	$	&	$	0.45	\pm	0.08	$	&	$	0.89	\pm	0.36	$	&	$	-5.34	\pm	12.49	$	&	$	65.15	\pm	12.29	$	&	$	0.58	\pm	0.18	$	&	$	-8.19	\pm	12.21	$	&	$	68.13	\pm	12.60	$	\\
49	&	PSZ2~G066.41+27.03	&	$	0.5699	$	&	$	0.79	\pm	0.16	$	&	$	1.76	\pm	0.73	$	&	$	-37.37	\pm	11.95	$	&	$	100.92	\pm	13.20	$	&	$	1.00	\pm	0.24	$	&	$	-34.28	\pm	11.21	$	&	$	97.77	\pm	11.89	$	\\
50	&	PSZ2~G144.83+25.11	&	$	0.584	$	&	$	0.61	\pm	0.07	$	&	$	1.34	\pm	0.45	$	&	$	1.55	\pm	9.00	$	&	$	-3.86	\pm	8.95	$	&	$	0.89	\pm	0.17	$	&	$	3.09	\pm	8.57	$	&	$	-2.97	\pm	8.79	$	\\
51	&	PSZ2~G045.87+57.70	&	$	0.611	$	&	$	0.41	\pm	0.12	$	&	$	0.93	\pm	0.46	$	&	$	20.59	\pm	17.97	$	&	$	16.79	\pm	15.76	$	&	$	0.52	\pm	0.16	$	&	$	16.61	\pm	16.65	$	&	$	20.54	\pm	14.20	$	\\
52	&	PSZ2~G108.27+48.66	&	$	0.674	$	&	$	0.40	\pm	0.08	$	&	$	0.55	\pm	0.20	$	&	$	8.45	\pm	11.83	$	&	$	35.26	\pm	11.93	$	&	$	0.42	\pm	0.12	$	&	$	9.91	\pm	12.03	$	&	$	35.53	\pm	11.69	$	\\
53	&	PSZ2~G086.93+53.18	&	$	0.6752	$	&	$	0.43	\pm	0.10	$	&	$	1.28	\pm	0.57	$	&	$	-40.06	\pm	16.39	$	&	$	30.84	\pm	12.08	$	&	$	0.59	\pm	0.15	$	&	$	-44.92	\pm	15.26	$	&	$	29.36	\pm	11.53	$	\\
54	&	PSZ2~G141.77+14.19	&	$	0.83	$	&	$	0.45	\pm	0.06	$	&	$	0.56	\pm	0.17	$	&	$	-3.40	\pm	8.77	$	&	$	-18.18	\pm	9.36	$	&	$	0.47	\pm	0.11	$	&	$	-4.11	\pm	8.78	$	&	$	-18.97	\pm	9.37	$	\\

\hline

\end{longtable}
\end{center}

\begin{center}
\begin{longtable}{lllllllllll}
\caption{Summary of model comparison statistics for final sample of 54 clusters. The \textit{Planck} IDs are omitted but are the same as in Table~\ref{tab:results1}.}\label{tab:results2} \\

\hline 
\multicolumn{1}{c}{Row} & \multicolumn{1}{c}{$z$} & \multicolumn{1}{c}{$d_{\rm EMD}(\mathcal{P}_{\rm PM}, \mathcal{P}_{\rm OM\, I})$} & \multicolumn{1}{c}{$d_{\rm EMD}(\mathcal{P}_{\rm OM\, II}, \mathcal{P}_{\rm OM\, I})$} & \multicolumn{1}{c}{$d_{\rm EMD}(\mathcal{P}_{\rm PM}, \mathcal{P}_{\rm OM\, II})$} & \multicolumn{1}{c}{$\ln (\mathcal{Z}_{\rm PM} / \mathcal{Z}_{\rm null})$} & \multicolumn{1}{c}{$\ln (\mathcal{Z}_{\rm OM \, I} / \mathcal{Z}_{\rm null})$} & \multicolumn{1}{c}{$\ln (\mathcal{Z}_{\rm OM \, II} / \mathcal{Z}_{\rm null})$} & \multicolumn{1}{c}{$\ln (\mathcal{Z}_{\rm PM} / \mathcal{Z}_{\rm OM \, I})$} & \multicolumn{1}{c}{$\ln (\mathcal{Z}_{\rm OM \, II} / \mathcal{Z}_{\rm OM \, I})$} & \multicolumn{1}{c}{$\ln (\mathcal{Z}_{\rm PM} / \mathcal{Z}_{\rm OM \, II})$} \\ \hline 
\endfirsthead

\multicolumn{11}{c}%
{{\tablename\ \thetable{} -- continued from previous page}} \\
\hline 
\multicolumn{1}{c}{Row} & \multicolumn{1}{c}{$z$} & \multicolumn{1}{c}{$d_{\rm EMD}(\mathcal{P}_{\rm PM}, \mathcal{P}_{\rm OM\, I})$} & \multicolumn{1}{c}{$d_{\rm EMD}(\mathcal{P}_{\rm OM\, II}, \mathcal{P}_{\rm OM\, I})$} & \multicolumn{1}{c}{$d_{\rm EMD}(\mathcal{P}_{\rm PM}, \mathcal{P}_{\rm OM\, II})$} & \multicolumn{1}{c}{$\ln (\mathcal{Z}_{\rm PM} / \mathcal{Z}_{\rm null})$} & \multicolumn{1}{c}{$\ln (\mathcal{Z}_{\rm OM \, I} / \mathcal{Z}_{\rm null})$} & \multicolumn{1}{c}{$\ln (\mathcal{Z}_{\rm OM \, II} / \mathcal{Z}_{\rm null})$} & \multicolumn{1}{c}{$\ln (\mathcal{Z}_{\rm PM} / \mathcal{Z}_{\rm OM \, I})$} & \multicolumn{1}{c}{$\ln (\mathcal{Z}_{\rm OM \, II} / \mathcal{Z}_{\rm OM \, I})$} & \multicolumn{1}{c}{$\ln (\mathcal{Z}_{\rm PM} / \mathcal{Z}_{\rm OM \, II})$} \\ 
\hline 
\endhead

\tabulinesep=_1mm
\extrarowsep=1mm
\LTcapwidth=\textwidth

1	&	$	0.0894	$	&	$	0.222	$	&	$	0.514	$	&	$	0.297	$	&	$	33.90	\pm	0.16	$	&	$	29.17	\pm	0.16	$	&	$	33.38	\pm	0.16	$	&	$	4.73	\pm	0.23	$	&	$	4.21	\pm	0.23	$	&	$	0.52	\pm	0.22	$	\\
2	&	$	0.113	$	&	$	0.152	$	&	$	0.091	$	&	$	0.093	$	&	$	30.94	\pm	0.17	$	&	$	31.06	\pm	0.17	$	&	$	30.01	\pm	0.17	$	&	$	-0.12	\pm	0.24	$	&	$	-1.05	\pm	0.24	$	&	$	0.93	\pm	0.24	$	\\
3	&	$	0.12	$	&	$	0.083	$	&	$	0.123	$	&	$	0.189	$	&	$	10.40	\pm	0.13	$	&	$	10.54	\pm	0.13	$	&	$	10.00	\pm	0.14	$	&	$	-0.14	\pm	0.19	$	&	$	-0.53	\pm	0.19	$	&	$	0.39	\pm	0.19	$	\\
4	&	$	0.137	$	&	$	0.132	$	&	$	0.115	$	&	$	0.051	$	&	$	1.71	\pm	0.17	$	&	$	3.41	\pm	0.17	$	&	$	1.76	\pm	0.17	$	&	$	-1.70	\pm	0.24	$	&	$	-1.65	\pm	0.24	$	&	$	-0.05	\pm	0.24	$	\\
5	&	$	0.1427	$	&	$	0.170	$	&	$	0.033	$	&	$	0.138	$	&	$	23.01	\pm	0.15	$	&	$	24.85	\pm	0.15	$	&	$	23.50	\pm	0.15	$	&	$	-1.84	\pm	0.21	$	&	$	-1.35	\pm	0.21	$	&	$	-0.49	\pm	0.21	$	\\
6	&	$	0.144	$	&	$	0.210	$	&	$	0.045	$	&	$	0.165	$	&	$	13.68	\pm	0.13	$	&	$	17.82	\pm	0.13	$	&	$	15.56	\pm	0.14	$	&	$	-4.14	\pm	0.18	$	&	$	-2.26	\pm	0.19	$	&	$	-1.88	\pm	0.19	$	\\
7	&	$	0.147	$	&	$	0.140	$	&	$	0.014	$	&	$	0.126	$	&	$	32.94	\pm	0.12	$	&	$	34.76	\pm	0.12	$	&	$	33.50	\pm	0.12	$	&	$	-1.82	\pm	0.17	$	&	$	-1.26	\pm	0.17	$	&	$	-0.56	\pm	0.17	$	\\
8	&	$	0.164	$	&	$	0.065	$	&	$	0.026	$	&	$	0.069	$	&	$	9.61	\pm	0.08	$	&	$	10.32	\pm	0.08	$	&	$	9.10	\pm	0.08	$	&	$	-0.71	\pm	0.11	$	&	$	-1.23	\pm	0.11	$	&	$	0.51	\pm	0.12	$	\\
9	&	$	0.1709	$	&	$	0.049	$	&	$	0.082	$	&	$	0.087	$	&	$	33.10	\pm	0.16	$	&	$	33.00	\pm	0.16	$	&	$	32.62	\pm	0.16	$	&	$	0.10	\pm	0.22	$	&	$	-0.37	\pm	0.22	$	&	$	0.47	\pm	0.23	$	\\
10	&	$	0.171	$	&	$	0.058	$	&	$	0.012	$	&	$	0.058	$	&	$	7.73	\pm	0.15	$	&	$	8.46	\pm	0.15	$	&	$	7.08	\pm	0.15	$	&	$	-0.73	\pm	0.21	$	&	$	-1.38	\pm	0.21	$	&	$	0.65	\pm	0.21	$	\\
11	&	$	0.1712	$	&	$	0.135	$	&	$	0.022	$	&	$	0.114	$	&	$	26.98	\pm	0.10	$	&	$	28.19	\pm	0.10	$	&	$	27.08	\pm	0.11	$	&	$	-1.21	\pm	0.14	$	&	$	-1.11	\pm	0.15	$	&	$	-0.10	\pm	0.15	$	\\
12	&	$	0.174	$	&	$	0.132	$	&	$	0.029	$	&	$	0.107	$	&	$	3.67	\pm	0.11	$	&	$	4.53	\pm	0.11	$	&	$	3.56	\pm	0.11	$	&	$	-0.86	\pm	0.15	$	&	$	-0.97	\pm	0.16	$	&	$	0.11	\pm	0.16	$	\\
13	&	$	0.181	$	&	$	0.084	$	&	$	0.015	$	&	$	0.080	$	&	$	4.42	\pm	0.13	$	&	$	5.00	\pm	0.12	$	&	$	4.06	\pm	0.13	$	&	$	-0.58	\pm	0.18	$	&	$	-0.95	\pm	0.18	$	&	$	0.36	\pm	0.18	$	\\
14	&	$	0.183	$	&	$	0.068	$	&	$	0.010	$	&	$	0.063	$	&	$	5.57	\pm	0.13	$	&	$	6.52	\pm	0.13	$	&	$	5.35	\pm	0.13	$	&	$	-0.94	\pm	0.18	$	&	$	-1.16	\pm	0.18	$	&	$	0.22	\pm	0.19	$	\\
15	&	$	0.185	$	&	$	0.062	$	&	$	0.010	$	&	$	0.056	$	&	$	3.57	\pm	0.18	$	&	$	4.28	\pm	0.18	$	&	$	3.47	\pm	0.18	$	&	$	-0.71	\pm	0.25	$	&	$	-0.80	\pm	0.25	$	&	$	0.09	\pm	0.25	$	\\
16	&	$	0.21	$	&	$	0.094	$	&	$	0.026	$	&	$	0.076	$	&	$	7.98	\pm	0.14	$	&	$	8.67	\pm	0.14	$	&	$	7.51	\pm	0.14	$	&	$	-0.69	\pm	0.20	$	&	$	-1.15	\pm	0.20	$	&	$	0.46	\pm	0.20	$	\\
17	&	$	0.2138	$	&	$	0.143	$	&	$	0.051	$	&	$	0.094	$	&	$	4.68	\pm	0.18	$	&	$	5.67	\pm	0.18	$	&	$	4.38	\pm	0.18	$	&	$	-0.99	\pm	0.25	$	&	$	-1.29	\pm	0.25	$	&	$	0.30	\pm	0.25	$	\\
18	&	$	0.2172	$	&	$	0.072	$	&	$	0.006	$	&	$	0.069	$	&	$	27.82	\pm	0.12	$	&	$	28.93	\pm	0.12	$	&	$	27.64	\pm	0.13	$	&	$	-1.11	\pm	0.17	$	&	$	-1.29	\pm	0.17	$	&	$	0.18	\pm	0.18	$	\\
19	&	$	0.223	$	&	$	0.097	$	&	$	0.054	$	&	$	0.057	$	&	$	4.36	\pm	0.10	$	&	$	4.84	\pm	0.10	$	&	$	3.95	\pm	0.10	$	&	$	-0.48	\pm	0.14	$	&	$	-0.89	\pm	0.14	$	&	$	0.41	\pm	0.14	$	\\
20	&	$	0.224	$	&	$	0.049	$	&	$	0.009	$	&	$	0.051	$	&	$	16.34	\pm	0.13	$	&	$	17.23	\pm	0.13	$	&	$	15.79	\pm	0.13	$	&	$	-0.89	\pm	0.18	$	&	$	-1.44	\pm	0.19	$	&	$	0.55	\pm	0.19	$	\\
21	&	$	0.232	$	&	$	0.225	$	&	$	0.147	$	&	$	0.083	$	&	$	3.02	\pm	0.15	$	&	$	4.37	\pm	0.15	$	&	$	2.54	\pm	0.15	$	&	$	-1.35	\pm	0.21	$	&	$	-1.82	\pm	0.21	$	&	$	0.48	\pm	0.21	$	\\
22	&	$	0.25	$	&	$	0.136	$	&	$	0.071	$	&	$	0.070	$	&	$	3.03	\pm	0.15	$	&	$	3.96	\pm	0.15	$	&	$	2.26	\pm	0.15	$	&	$	-0.93	\pm	0.21	$	&	$	-1.70	\pm	0.21	$	&	$	0.77	\pm	0.21	$	\\
23	&	$	0.251	$	&	$	0.055	$	&	$	0.024	$	&	$	0.045	$	&	$	12.67	\pm	0.16	$	&	$	13.45	\pm	0.16	$	&	$	11.69	\pm	0.17	$	&	$	-0.78	\pm	0.23	$	&	$	-1.76	\pm	0.23	$	&	$	0.99	\pm	0.23	$	\\
24	&	$	0.254	$	&	$	0.180	$	&	$	0.110	$	&	$	0.076	$	&	$	3.80	\pm	0.11	$	&	$	5.27	\pm	0.11	$	&	$	3.86	\pm	0.11	$	&	$	-1.47	\pm	0.15	$	&	$	-1.41	\pm	0.15	$	&	$	-0.06	\pm	0.15	$	\\
25	&	$	0.26	$	&	$	0.041	$	&	$	0.009	$	&	$	0.040	$	&	$	13.18	\pm	0.17	$	&	$	13.79	\pm	0.16	$	&	$	12.32	\pm	0.17	$	&	$	-0.60	\pm	0.23	$	&	$	-1.46	\pm	0.23	$	&	$	0.86	\pm	0.23	$	\\
26	&	$	0.2671	$	&	$	0.043	$	&	$	0.012	$	&	$	0.051	$	&	$	28.23	\pm	0.14	$	&	$	29.05	\pm	0.14	$	&	$	27.67	\pm	0.14	$	&	$	-0.81	\pm	0.20	$	&	$	-1.38	\pm	0.20	$	&	$	0.56	\pm	0.20	$	\\
27	&	$	0.288	$	&	$	0.038	$	&	$	0.018	$	&	$	0.032	$	&	$	22.61	\pm	0.14	$	&	$	23.45	\pm	0.14	$	&	$	21.90	\pm	0.14	$	&	$	-0.85	\pm	0.19	$	&	$	-1.55	\pm	0.19	$	&	$	0.71	\pm	0.20	$	\\
28	&	$	0.29	$	&	$	0.045	$	&	$	0.034	$	&	$	0.046	$	&	$	9.72	\pm	0.18	$	&	$	10.64	\pm	0.18	$	&	$	9.42	\pm	0.18	$	&	$	-0.92	\pm	0.26	$	&	$	-1.23	\pm	0.26	$	&	$	0.31	\pm	0.26	$	\\
29	&	$	0.3	$	&	$	0.138	$	&	$	0.115	$	&	$	0.045	$	&	$	5.26	\pm	0.20	$	&	$	5.94	\pm	0.19	$	&	$	4.44	\pm	0.20	$	&	$	-0.68	\pm	0.28	$	&	$	-1.51	\pm	0.28	$	&	$	0.83	\pm	0.28	$	\\
30	&	$	0.3092	$	&	$	0.047	$	&	$	0.027	$	&	$	0.041	$	&	$	14.83	\pm	0.12	$	&	$	15.62	\pm	0.12	$	&	$	14.13	\pm	0.12	$	&	$	-0.80	\pm	0.17	$	&	$	-1.49	\pm	0.17	$	&	$	0.70	\pm	0.17	$	\\
31	&	$	0.3184	$	&	$	0.042	$	&	$	0.025	$	&	$	0.032	$	&	$	14.64	\pm	0.11	$	&	$	15.36	\pm	0.10	$	&	$	13.88	\pm	0.11	$	&	$	-0.72	\pm	0.15	$	&	$	-1.48	\pm	0.15	$	&	$	0.76	\pm	0.15	$	\\
32	&	$	0.333	$	&	$	0.071	$	&	$	0.058	$	&	$	0.036	$	&	$	9.30	\pm	0.15	$	&	$	9.89	\pm	0.15	$	&	$	8.59	\pm	0.15	$	&	$	-0.58	\pm	0.21	$	&	$	-1.30	\pm	0.21	$	&	$	0.72	\pm	0.21	$	\\
33	&	$	0.336	$	&	$	0.209	$	&	$	0.122	$	&	$	0.088	$	&	$	10.98	\pm	0.20	$	&	$	14.24	\pm	0.19	$	&	$	12.05	\pm	0.20	$	&	$	-3.26	\pm	0.28	$	&	$	-2.19	\pm	0.28	$	&	$	-1.07	\pm	0.28	$	\\
34	&	$	0.341	$	&	$	0.032	$	&	$	0.006	$	&	$	0.031	$	&	$	32.32	\pm	0.14	$	&	$	33.03	\pm	0.14	$	&	$	31.53	\pm	0.14	$	&	$	-0.71	\pm	0.20	$	&	$	-1.51	\pm	0.20	$	&	$	0.80	\pm	0.20	$	\\
35	&	$	0.353	$	&	$	0.036	$	&	$	0.016	$	&	$	0.045	$	&	$	20.74	\pm	0.16	$	&	$	21.70	\pm	0.15	$	&	$	20.26	\pm	0.16	$	&	$	-0.96	\pm	0.22	$	&	$	-1.44	\pm	0.22	$	&	$	0.48	\pm	0.22	$	\\
36	&	$	0.356	$	&	$	0.039	$	&	$	0.007	$	&	$	0.043	$	&	$	25.23	\pm	0.13	$	&	$	25.70	\pm	0.13	$	&	$	24.79	\pm	0.14	$	&	$	-0.47	\pm	0.19	$	&	$	-0.91	\pm	0.19	$	&	$	0.44	\pm	0.19	$	\\
37	&	$	0.366	$	&	$	0.037	$	&	$	0.018	$	&	$	0.027	$	&	$	11.84	\pm	0.13	$	&	$	12.47	\pm	0.13	$	&	$	11.00	\pm	0.13	$	&	$	-0.62	\pm	0.19	$	&	$	-1.47	\pm	0.19	$	&	$	0.84	\pm	0.19	$	\\
38	&	$	0.375	$	&	$	0.093	$	&	$	0.050	$	&	$	0.047	$	&	$	16.17	\pm	0.14	$	&	$	17.58	\pm	0.14	$	&	$	15.83	\pm	0.14	$	&	$	-1.41	\pm	0.20	$	&	$	-1.74	\pm	0.20	$	&	$	0.34	\pm	0.20	$	\\
39	&	$	0.4	$	&	$	0.023	$	&	$	0.013	$	&	$	0.027	$	&	$	3.36	\pm	0.15	$	&	$	2.77	\pm	0.15	$	&	$	2.75	\pm	0.15	$	&	$	0.59	\pm	0.22	$	&	$	-0.02	\pm	0.22	$	&	$	0.61	\pm	0.22	$	\\
40	&	$	0.412	$	&	$	0.040	$	&	$	0.015	$	&	$	0.054	$	&	$	26.82	\pm	0.16	$	&	$	27.58	\pm	0.16	$	&	$	26.56	\pm	0.16	$	&	$	-0.76	\pm	0.23	$	&	$	-1.01	\pm	0.23	$	&	$	0.26	\pm	0.23	$	\\
41	&	$	0.422	$	&	$	0.058	$	&	$	0.027	$	&	$	0.032	$	&	$	14.70	\pm	0.22	$	&	$	15.84	\pm	0.22	$	&	$	14.37	\pm	0.22	$	&	$	-1.14	\pm	0.31	$	&	$	-1.48	\pm	0.31	$	&	$	0.33	\pm	0.31	$	\\
42	&	$	0.426	$	&	$	0.126	$	&	$	0.106	$	&	$	0.030	$	&	$	4.48	\pm	0.15	$	&	$	4.89	\pm	0.14	$	&	$	4.24	\pm	0.15	$	&	$	-0.41	\pm	0.20	$	&	$	-0.66	\pm	0.20	$	&	$	0.24	\pm	0.21	$	\\
43	&	$	0.45	$	&	$	0.025	$	&	$	0.010	$	&	$	0.020	$	&	$	31.61	\pm	0.16	$	&	$	32.30	\pm	0.15	$	&	$	30.87	\pm	0.16	$	&	$	-0.69	\pm	0.22	$	&	$	-1.43	\pm	0.22	$	&	$	0.74	\pm	0.22	$	\\
44	&	$	0.47	$	&	$	0.032	$	&	$	0.023	$	&	$	0.041	$	&	$	8.28	\pm	0.10	$	&	$	8.74	\pm	0.10	$	&	$	8.14	\pm	0.11	$	&	$	-0.46	\pm	0.14	$	&	$	-0.60	\pm	0.15	$	&	$	0.14	\pm	0.15	$	\\
45	&	$	0.5131	$	&	$	0.133	$	&	$	0.078	$	&	$	0.055	$	&	$	23.66	\pm	0.14	$	&	$	27.24	\pm	0.13	$	&	$	24.82	\pm	0.14	$	&	$	-3.58	\pm	0.19	$	&	$	-2.42	\pm	0.19	$	&	$	-1.16	\pm	0.19	$	\\
46	&	$	0.5392	$	&	$	0.036	$	&	$	0.007	$	&	$	0.043	$	&	$	23.74	\pm	0.18	$	&	$	24.69	\pm	0.18	$	&	$	24.49	\pm	0.18	$	&	$	-0.95	\pm	0.25	$	&	$	-0.20	\pm	0.25	$	&	$	-0.75	\pm	0.25	$	\\
47	&	$	0.545	$	&	$	0.028	$	&	$	0.010	$	&	$	0.020	$	&	$	110.33	\pm	0.19	$	&	$	110.78	\pm	0.19	$	&	$	109.81	\pm	0.19	$	&	$	-0.45	\pm	0.26	$	&	$	-0.97	\pm	0.26	$	&	$	0.52	\pm	0.27	$	\\
48	&	$	0.5586	$	&	$	0.064	$	&	$	0.041	$	&	$	0.027	$	&	$	21.75	\pm	0.20	$	&	$	22.86	\pm	0.20	$	&	$	21.54	\pm	0.20	$	&	$	-1.11	\pm	0.28	$	&	$	-1.31	\pm	0.28	$	&	$	0.21	\pm	0.28	$	\\
49	&	$	0.5699	$	&	$	0.101	$	&	$	0.071	$	&	$	0.031	$	&	$	14.90	\pm	0.17	$	&	$	16.67	\pm	0.17	$	&	$	14.44	\pm	0.17	$	&	$	-1.77	\pm	0.24	$	&	$	-2.23	\pm	0.24	$	&	$	0.46	\pm	0.24	$	\\
50	&	$	0.584	$	&	$	0.080	$	&	$	0.041	$	&	$	0.039	$	&	$	43.03	\pm	0.17	$	&	$	45.57	\pm	0.17	$	&	$	43.52	\pm	0.17	$	&	$	-2.54	\pm	0.24	$	&	$	-2.05	\pm	0.24	$	&	$	-0.49	\pm	0.25	$	\\
51	&	$	0.611	$	&	$	0.112	$	&	$	0.079	$	&	$	0.035	$	&	$	8.60	\pm	0.14	$	&	$	10.46	\pm	0.14	$	&	$	8.54	\pm	0.14	$	&	$	-1.86	\pm	0.20	$	&	$	-1.92	\pm	0.20	$	&	$	0.06	\pm	0.20	$	\\
52	&	$	0.674	$	&	$	0.032	$	&	$	0.026	$	&	$	0.015	$	&	$	13.43	\pm	0.16	$	&	$	13.61	\pm	0.16	$	&	$	12.69	\pm	0.16	$	&	$	-0.18	\pm	0.23	$	&	$	-0.92	\pm	0.23	$	&	$	0.74	\pm	0.23	$	\\
53	&	$	0.6752	$	&	$	0.126	$	&	$	0.090	$	&	$	0.037	$	&	$	13.17	\pm	0.13	$	&	$	15.96	\pm	0.13	$	&	$	13.48	\pm	0.14	$	&	$	-2.79	\pm	0.19	$	&	$	-2.48	\pm	0.19	$	&	$	-0.32	\pm	0.19	$	\\
54	&	$	0.83	$	&	$	0.020	$	&	$	0.014	$	&	$	0.013	$	&	$	35.45	\pm	0.12	$	&	$	35.38	\pm	0.11	$	&	$	34.60	\pm	0.12	$	&	$	0.07	\pm	0.16	$	&	$	-0.78	\pm	0.17	$	&	$	0.85	\pm	0.17	$	\\

\hline

\end{longtable}
\end{center}

\bsp	
\label{lastpage}
\end{landscape}
\restoregeometry
\end{document}